\documentclass[a4paper,12pt]{article}
\usepackage[utf8]{inputenc}
\usepackage[T1]{fontenc}
\usepackage[english]{babel}
\usepackage{hyperref}
\usepackage{natbib}

\usepackage{color} % Allow text colors
\usepackage{amsmath, amscd, amsfonts}
\usepackage{graphicx}
\usepackage{subfigure}

\usepackage{graphics}
\usepackage{amssymb}
\usepackage{epsfig}
\usepackage{rotating}
\usepackage{subfigure}
\usepackage{changepage}
\usepackage{parskip}
\usepackage{listing}
\usepackage{eso-pic}
\usepackage{mathrsfs}
\usepackage{threeparttable}
\usepackage{setspace}
\usepackage{times}
\usepackage{multirow}
\usepackage{array}
\usepackage{tabularx}
\usepackage{color}
\usepackage{longtable}
\usepackage{setspace}
\usepackage{multirow}
\usepackage{array}
\usepackage{booktabs,caption}
\usepackage{threeparttable}
\newcolumntype{P}[1]{>{\centering\arraybackslash}p{#1}}
\newcommand{\mathbbm}[1]{\text{\usefont{U}{bbm}{m}{n}#1}}

\newcommand{\E}{\mathop{\mbox{\sf E}}}
\newcommand{\N}{\mathop{\mbox{\sf N}}}

\newcommand{\Cov}{{\sf Cov}}

\def\defeq{\stackrel{\mathrm{def}}{=}}
\advance\textheight by \topskip
\everymath{\fontdimen16\textfont2=2.7pt\fontdimen17\textfont2=2.7pt}

\begin{document}
%\date{20180114.0900}
%\pagestyle{headings}
\pagestyle{plain}

\title{\textbf{Pricing cryptocurrency options}}

\author{ Ai Jun Hou\thanks{Stockholm Business School, Stockholm University.}, Weining Wang\thanks{Department of Economics and Related Studies, University of York.}, Cathy Y.H. Chen\thanks{Adam Smith Business School, University of Glasgow.}, Wolfgang Karl H\"ardle\thanks{School of Business and Economics, Ladislaus von Bortkiewicz Chair of Statistics, Humboldt-Universit\"{a}t zu Berlin. Sim Kee Boon Institute, Singapore Management University; Faculty of Mathematics and Physics, Charles University in Prague. }}

\maketitle \doublespace
\begin{abstract}
Cryptocurrencies, especially Bitcoin (BTC), which comprise a new digital asset class, have drawn extraordinary worldwide attention. 
The characteristics of the cryptocurrency/BTC include a high level of speculation, extreme volatility and price discontinuity. We propose a pricing mechanism based on a stochastic volatility with a correlated jump (SVCJ) model and compare it to a flexible co-jump model by \cite{bandi2016price}. The estimation results of both models confirm the impact of jumps and co-jumps on options obtained via simulation and an analysis of the implied volatility curve. We show that a sizeable proportion of price jumps are significantly and contemporaneously anti-correlated with jumps in volatility. Our study comprises pioneering research on pricing BTC options. We show how the proposed pricing mechanism underlines the importance of jumps in cryptocurrency markets. 
\end{abstract}

Acknowledgement: We thank \cite{bandi2016price} for providing us with the codes used to implement the BR co-jumps model in this study. We acknowledge the helpful comments from the editor and two anonymous referees. We thank Xiaohao Ji for help with data processing. This research is supported by the Deutsche Forschungsgemeinschaft through the International Research Training Group 1792 "High Dimensional Nonstationary Time Series".  (http://irtg1792.hu-berlin.de).  In addition, it has been funded by the Natural Science Foundation of China (fund number 71528008). Ai Jun Hou acknowledges the financial support from the Jan Wallender and Tom Hedelius Foundation of Handelsbanken (P2019-0264). H\"ardle acknowledges the financial support from the Czech Science Foundation and the Yushan Scholar Program. \\
\textit{This is a post-peer-review, pre-copyedit version of an article published in the Journal of Financial Econometrics. The final authenticated version is available online at: }\url{http://dx.doi.org/10.1093/jjfinec/nbaa006}.

\section{Introduction}

Bitcoin (BTC), the network-based decentralized digital currency and payment system, has garnered worldwide attention and interest since it was first introduced in 2009. The rapidly growing research related to BTC shows a prominent role of this new digital asset class in contemporary financial markets.\footnote{ see e.g., \cite{becker2013can},\cite{segendorf2014bitcoin},\cite{dwyer2015economics}, also studies on economics \citep{kroll2013economics}, alternative monetary systems (\cite{RB2014} and \cite{Weber2016}) and financial stability \citep{Ali2014, BC2014,ECB2015}. An analysis of the legal issues involved in using Bitcoin can be found in \cite{Murphy2015}.} Several studies have suggested econometric methods to model the dynamics of BTC prices, including cross-sectional regression models involving the major traded cryptocurrencies and also multivariate time-series models.\footnote{For example, \cite{hayes2017cryptocurrency} performs a regression using a cross-section dataset consisting of 66 traded digital currencies to understand the price driver of cryptocurrencies. \cite{Kris2013} proposes a bivariate Vector-AutoRegression (VAR) model for the weekly log returns of Bitcoin prices. \cite{BST2015} investigates the long and short-run relationships between BTC prices and other related variables using an autoregressive distributed lag model.} \cite{scaillet2017high} show that jumps are much more frequent in the BTC market than, for example, in the US equity market (see e.g., \cite{bajgrowicz2015jumps}, \cite{eraker2004stock}, \cite{bandi2016price} and among others).{These earlier studies suggest that jumps should be considered when modeling BTC prices.} %that among the characteristics adhere to BTC, jump appears to be a landmark. 

However, research on the BTC derivative markets is still limited despite the rapidly growing availability of BTC futures and options traded on an unregulated exchange platform (i.e., \href{www.deribit.com}{Deribit}). Especially, the CME (Chicago Mercantile Exchange) Group, the world's leading derivatives marketplace, launched BTC futures based on the CME CF Bitcoin Reference Rate (BRR) on 18 December 2017. The limited research on pricing and hedging Bitcoin derivatives is partly attributed to the fact that pricing BTC derivatives (e.g., options) encounters econometric challenges from the extraordinary occurrence of jumps as this market is unregulated, lacks of central settlement and is highly speculation-driven. This calls for a more flexible model to capture the sudden jumps appearing in both the returns and variance processes.

In this paper, we contribute to the existing literature by exploring the stochastic and econometric properties of BTC dynamics and {then pricing the BTC options based on these properties.} 
The investigation is carried out by using the most advanced stochastic volatility models, i.e., the stochastic volatility with the correlated jump (SVCJ) model of \cite{duffie2000transform} and the stochastic volatility with the possible non-linearity structure of \cite{bandi2016price}(BR hereafter).  The employed SVCJ model incorporates jumps in both returns and the stochastic volatility process, while the BR model captures the possible {non-linearity of} return and variance processes and characterizes a non-affine structure. We aim to provide a theoretical foundation for the future development of derivative markets on cryptocurrencies.

Numerous empirical studies have applied the SVCJ model in different markets. For example, \cite{eraker2003impact} and \cite{eraker2004stock} use the SVCJ model to describe equity market returns and estimate equity option pricing. They find strong evidence of jumps in returns and volatility in the US equity market. %\cite{cosma2016early} develop a methodology to estimate a jump-diffusion model to price American options or other path-dependent options. To show the importance of modeling jumps in the BTC returns and variance dynamics, 
We further compare the SVCJ {estimates} to the simplified versions such as \cite{bates2000post} (SVJ hereafter) and the stochastic volatility (SV) model.  

%Furthermore, %studies such as \cite{Bakshi2006} and \cite{CD2011}, among others, propose a non-affine structure for the volatility process.  
%there is also a strand of literature in which the nonparametric model setup is used to analyze whether jumps in returns and variance are important model components \citep{AIT2010, BarnS2006}. These studies suggest that jumps and non-affine structures are potentially important model components.
%Studies such as \cite{Bakshi2006} and \cite{CD2011}, among others, propose a non-affine structure for the volatility process.  There is also a strand of literature in which the nonparametric model setup is used to analyze whether jumps in returns and variance are important model components \citep{AIT2010, BarnS2006}. These studies suggest that jumps and non-affine structures are potentially important model components.

For a purpose of robustness check, we compare our results with {those from the BR model}. \cite{bandi2016price} propose a price and variance co-jump model that generalizes the SVCJ model to capture the possible nonlinearity in the parameters of the returns and variance processes. The BR model characterizes independent and correlated jumps and allows for a nonparametric parameter structure, and estimates the parameters by using high-frequency data. We also apply this model to the dynamics of BTC. We base {our} option pricing on {an experimental simulation} where the parameters used to execute a simulation are from the SVCJ and BR model, respectively. 

%We may find the traded options data e.g. from \href{https://www.deribit.com/}{Deribit}, a leading Bitcoin future \& options exchange platform for plain vanilla European options, however, this trading platform lacks supervision from the authorities and clear regulations on clearing and settlement, the trading activities remain skeptical and the usage of this dataset is debatable. We compromise this restriction and broadly make use of model simulated option prices and even intraday BTC prices in our investigations. 

%We use the daily price from 31/07/2014 to 29/09/2017 to estimate the SVCJ model. We first calculate returns ($R_t$) as the log first difference between BTC prices, i.e. 
%\begin{equation*}
  %  $R_t=log(P_t)-ln(P_{t-1})$, where $P_t$ is the prices of BTC at time $t$.
%\end{equation*}
%Then we use returns to estimate the SVCJ model. The BR model is based on the intra-daily data from 31/07/2014 to 29/07/2017. All daily and intra-daily data are downloaded from Bloomberg. We first estimate the affine models (SVCJ, SVJ and SV) and nonaffine (BR) model parameters, then continue to analyze the option property by examining the simulated option prices and the Black Scholes model implied volatility. 

We summarize our main empirical findings as follows.
\textit{First}, as in the existing literature, the results from the SVCJ and BR models indicate that jumps are present in the returns and variance processes and adding jumps to the returns and volatility improves the goodness of fit.
\textit{Second}, in contrast to existing studies that commonly report a negative leverage effect, we find that the correlation between the return and volatility is significantly positive in the SVCJ model. However, we cannot find significant negative relations between risk and return in the BR model.  This implies that a rise of price is not associated with a decrease in volatility, which is %in favor of the notion of 
{consistent with} the "inverse leverage effect" found in the commodity markets {\citep{Schwartz2009SV}}. %The positive relationship between risk and volatility is reflected in the option prices simulated based on the parameters estimated from the SVCJ model. 

\textit{Third}, we find that the jump size in the return and variance of BTC is anti-correlated. The parameter estimates of the jump size ($\rho_j$) from both the SVCJ and BR models are negative (though the SVCJ estimate is insignificant).
It is worth noting that the correlation between the price jump size and the volatility jump size turns out to be significant with a negative coefficient with high-frequency data, while tending to be insignificant for the SVCJ fitting using daily prices. This finding is in line with existing studies of the stock market from \cite{eraker2004stock}, \cite{duffie2000transform} and \cite{bandi2016price}, among others. For example,  \cite{bandi2016price} report an anti-correlation with the nonaffine structure. \cite{eraker2004stock} finds a negative correlation between jump size only when augmenting return data with options data, and the negative correlation between co-jump size being identified in the implied volatility smirk. 
Using high-frequency data, \cite{Jocob2009} and \cite{TT2010} also report that the large jump size of prices and volatility are strongly anti-correlated. 

%\textit{Fourth}, we find that the affine-based model performs as good as the non-affine specification, and the estimated parameters from two specifications are quite comparable with few exceptions. It is worth noting that the correlation between the price jump size and the volatility jump size turns to be significant with a negative coefficient with high frequency data, while tend to be not significant for the SVCJ fitting using daily price. 

\textit{Finally,} we observe that the option price level is prominently dominated by the level of volatility and therefore overwhelmingly affected by jumps in the volatility processes. The results from the plots of implied volatility (IV) indicate that adding jumps in the return increases the slope of the IV curves. The greater steepness of the IV curve can be strengthened by the presence of jumps in volatility. The presence of co-jumps enlarges the IV smile further. {As evidenced from the IVs curve, options with a short time to maturity are more sensitive to jumps and co-jumps}. To fulfill a hedge or speculation need from institutional investors, we replicate the entire analysis for the CRyptocurrency IndeX (CRIX), a market portfolio comprising leading cryptocurrencies (see more detail in {\tt www.thecrix.de}). A recent volatility index, VCRIX, created by \cite{kim2019vcrix} also shows the evidence of jumps in CRIX.

To summarize our contributions, this study is the first paper to extensively investigate the stochastic and econometric properties of BTC and incorporate these properties in the BTC options pricing. Our results have practical relevance in terms of  model selection for characterizing the BTC dynamics. We document the necessity of incorporating jumps in the returns and volatility processes of BTC, and we find that jumps play a critical role in the option prices. Our approach is readily applicable to pricing BTC options in reality. Our results are also important for policymakers to design appropriate regulations for trading BTC derivatives and for institutional investors to launch effective risk management and efficient portfolio strategies.

The paper is organized as follows. Section \ref{dyna} {briefly introduces the BTC market}. Section \ref{sec:affine} studies the BTC return and variance dynamics with the SV, SVJ and SVCJ models.  Fitting of the BR model is investigated in Section \ref{sec:nonAffine}. Section \ref{sec:opt} implements the option-pricing exercises. Section \ref{sec:crix} documents an examination of the CRIX, while Section \ref{sec:con} concludes the study. A few preliminary econometrics analysis and estimation results for the CRIX are in the Appendix. The codes for this research can be found in {\tt www.quantlet.de}.

% whereas it is overwhelmingly negative in the stock market.  This contradiction may be attributed to the predominance between positive and negative price jumps. Since the CC markets experienced more often positive jumps than negative jumps, it is not surprising with this disjunctive result. 

%even with controlled systematic biases. These systematic biases are typically illustrated by the smile in implied volatilities extracted from a cross section of options, sorted by the moneyness. 

%\input{Introduction_May13.tex}
\section{The BTC dynamic}\label{dyna}
We start by briefly introducing BTC. BTC was the first open source distributed cryptocurrency released in 2009, after it was introduced in a paper “Bitcoin: A Peer-to-Peer Electronic Cash System” by a developer under the pseudonym Satoshi Nakamoto. It is a digital, decentralized, partially anonymous currency, not backed by any government or other legal entity. %Bitcoins may be transacted with other bitcoins with the help of peer-to-peer technology carried out by the network. 
%It has no physical counterpart, but
%merely arbitrary (divisible) units that exist on this network. 
The system has a pre-programmed money supply that grows at a decreasing rate until reaching a fixed limit. Since all is based on open source, the design and control is open for all. Traditional currencies are managed by a central bank, while BTCs are not regulated by any authority; instead, they are maintained by a decentralized community. The transactions of bitcoins are recorded in the ledgers (known as the blockchain), which is maintained by a
network of computers (called 'miners’). 
% Miners maintain consensus in the blockchain through solving difficult mathematical problems, and are rewarded with bitcoins and optional (voluntary) transaction fees. The additional rewarded bitcoins are the mechanism that increases the bitcoin
% money supply. 
Since bitcoin is not a country-specific currency, international payments can be carried out more economically and efficiently. 
% The CRyptocurrency IndeX \includegraphics[scale=0.037]{figures/crix1} developed by \cite{hardle2015crix} provides a market measure which consists of a selection of representative cryptos. Through the exceptional channel of an ICO, a CC startup can bypass the rigorous and regulated capital-raising process required by venture capitalists or banks. The appendix lists the top 10 constituents used to construct the CRIX index. The mechanism of selecting CRIX constituents is explained in \cite{trimborn2018}.

\begin{figure}[htp]
\caption {BTC Prices and Returns}\label{fig:BTCprice}
\begin{center}
\begin{tabular}{cc}
\includegraphics[scale=0.4]{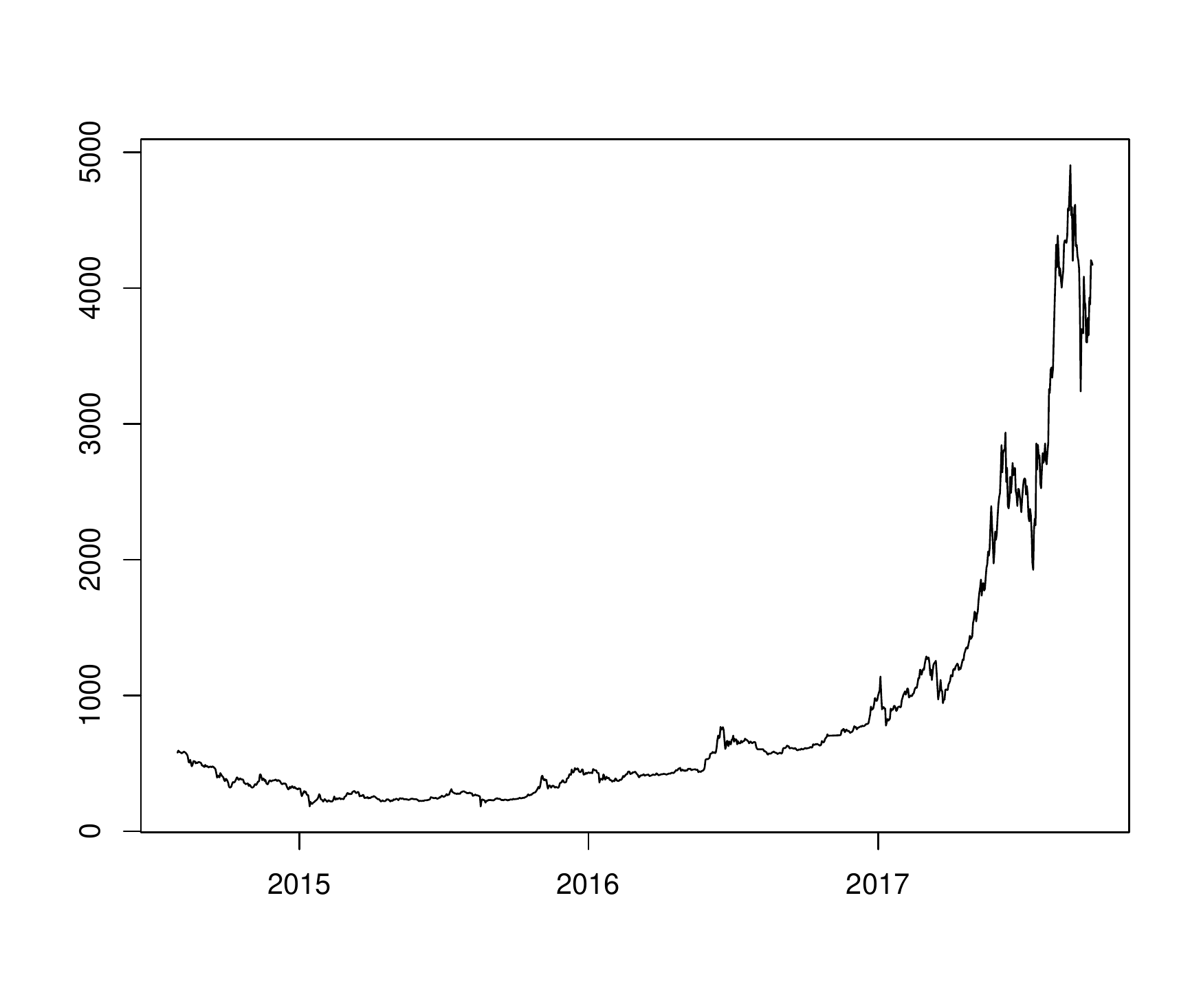} &
\includegraphics[scale=0.4]{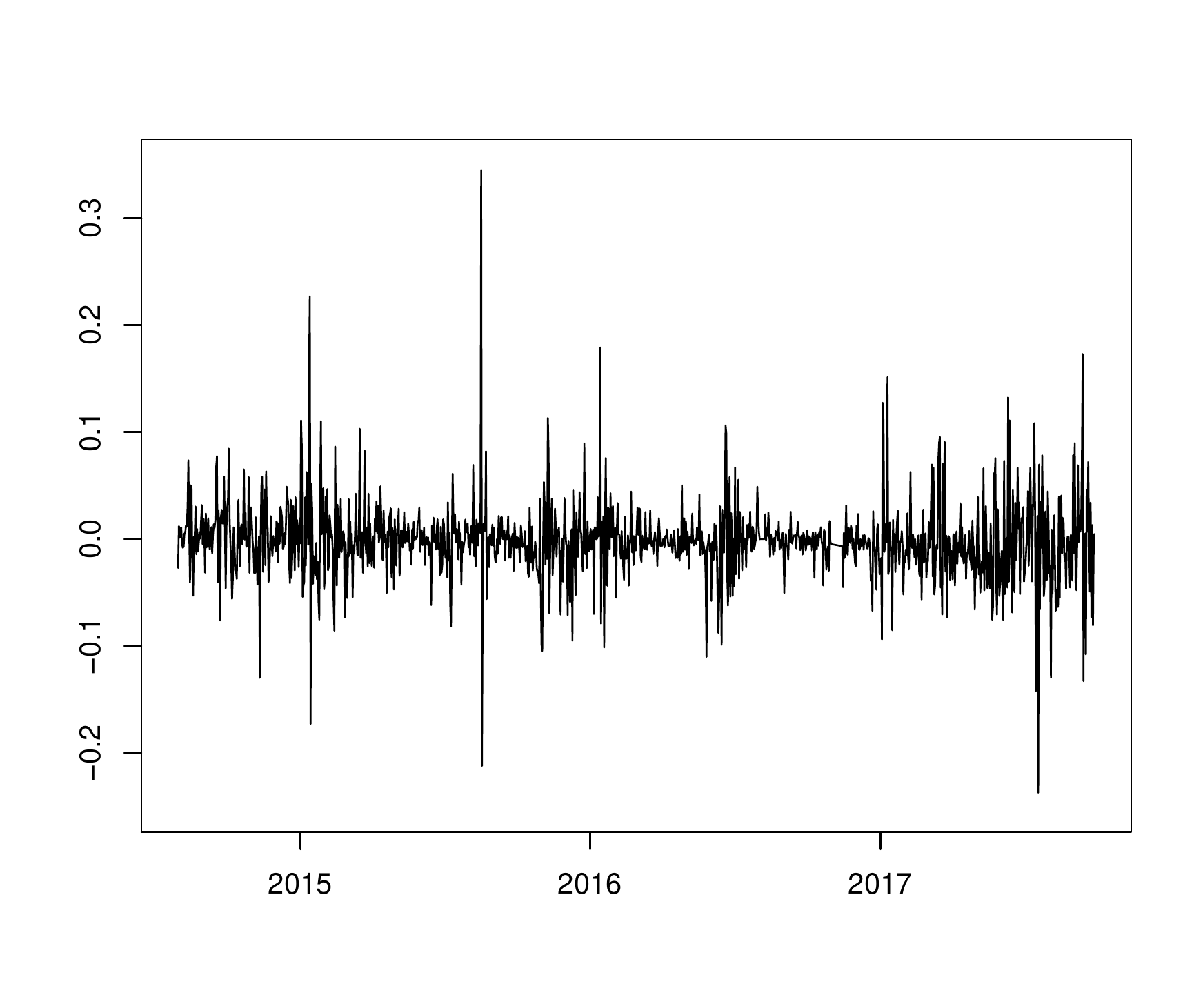}  \\
\end{tabular}
\end{center}
\footnotesize
\textit{Notes}: This figure graphs the BTC daily price (left panel) from 01/08/2014 to 29/09/2017 and BTC returns (right panel). The returns ($R_t$) are calculated as
%\begin{equation*}
    $R_t=\log(P_t)-\log(P_{t-1})$, where $P_t$ is the BTC price at time $t$.
%\end{equation*}
\end{figure}

Our empirical analyses are carried out based on both daily closing (SVCJ model) prices and five minutes intra-daily (BR model) prices. The data cover the period from 1 August 2014 to 29 September 2017 and are collected from Bloomberg. 
The dynamics of BTC daily prices (left panel) and BTC returns (right panel) are depicted in Figure \ref{fig:BTCprice}. It shows that the BTC return is clearly more volatile than the stock return, along with more frequent jumps or the scattered volatility spikes. Bitcoin’s price spent most of the year 2015 relatively stable. The BTC price in the first four months of 2016 was in the range of 400-460 USD. It moved upward dramatically after 2016 and increased to almost 5000 USD by the end of our sample period in 2017. At the time of the writing of this paper, the BTC market capitalization is more than USD 7 billions (source: Coinmarketcap 2017).

Both the BTC prices and returns react to big events in the BTC market.  A dramatic surge observed after March 2017 was due to the widespread interest in cryptocurrencies (CCs). The subsequent drop in June 2017 was caused by a sequence of political interventions. Several governmental announcements of bans on initial coin offerings (ICOs) have spurred intensive movements on CC markets. For example, the Chinese SEC (Securities and Exchange Commission) { denied} permission for a bitcoin ETF {on} March 10, {2017}; and Bitcoin crashed down after China banned initial coin offerings on September 4, 2017. 
 The large upward movements in BTC prices caused the returns of BTC displaying extremely high volatility and with scattered spikes/jumps. Several large jumps triggered by a series of big events in the BTC market can be detected from the returns series, see also \cite{kim2019vcrix}. We have implemented a number of time series models to the BTC returns and the results are shown in {Appendix \ref{sec:arima} and Appendix \ref{sec:garch}} . We find that the standard set of stationary models, such as ARIMA and GARCH, cannot fit the BTC returns well due to the presence of jumps.
 %These prices are far from stationary, with extremely high volatility and scattered spikes. 
%The plot of returns shows that volatility is time-varying and clustering. Therefore we start to fit the returns of BTC with the GRACH model. 

\section{SVCJ: affine specification} \label{sec:affine}
In this section, we estimate the SVCJ model using BTC prices. We begin with a simple SVCJ jump specification, and switch to the BR model in  Section \ref{sec:nonAffine}. We focus the analysis on BTC and then {introduce} CRIX in Section \ref{sec:crix}.

\subsection{Models}
In order to estimate the BTC dynamics with the SV and SVCJ models regarding returns and volatility, we employ the continuous time model of \cite{duffie2000transform} that encompasses the standard jump diffusion and the SV with jumps in returns only (SVJ) model of \cite{bates1996jumps}.  More precisely, let $\{S_t\}$ be
the price process, $\{d\log S_t\}$ the log returns and $\{V_t\}$ be the volatility process. The SVCJ dynamics are as follows:
\begin{eqnarray}
&d\log S_t  =  \mu dt+\sqrt{V_t}d W_t^{(S)}+Z_t^y d N_t\label{ss}\\
&dV_t  =
\kappa(\theta-V_t) dt+\sigma_V\sqrt{V_t}d W_t^{(V)}+Z_t^v dN_t\label{vv}\\
&\Cov(dW_t^{(S)},dW_t^{(V)})=\rho dt\\
&\operatorname{P} (dN_t=1)=\lambda dt.
\end{eqnarray}
Like in the Cox-Ingersoll-Ross model,
$\kappa$ and $\theta$ are the mean reversion rate and mean
reversion level, respectively.  $W^{(S)}$ and $W^{(V)}$ are two
correlated standard Brownian motions with correlation denoted as
$\rho$. $N_t$ is a pure jump process with {a}
constant mean jump-arrival rate ${\lambda}$. The random jump
sizes are $Z_t^y$ and $Z_t^v$. Since the jump-driving Poisson process is
the same in both (\ref{ss}), (\ref{vv}), the jump sizes can be
correlated.  The random jump size $Z_t^y$ conditional on $Z_t^v$, is assumed
to have a Gaussian distribution with a mean of $\mu_y+\rho_jZ_t^v$ and
standard deviation set to  $\sigma_y$.  The jump in volatility $Z_t^v$ is
assumed to follow an exponential distribution with mean $\mu_v$:
\begin{eqnarray}
Z_t^y|Z_t^v\sim\; \operatorname{N}(\mu_y+\rho_j Z_t^v, \sigma^2_y); \;\;\;
Z_t^v\sim\;\exp(\mu_v). \label{jumpdist}
\end{eqnarray}
The correlation $ \rho $ between the diffusion
terms is introduced to capture the possible leverage effects between returns
and volatility. The jumps may be correlated as well. The correlation term $ \rho_j  $ takes care of that. 
The SV process $\sqrt{V}_t$ is modelled as a square root
process. With no jumps in the volatility, the parameter $\theta$
is the long-run mean of $V_t$, and the process reverts to this level at
a speed governed by the parameter $\kappa$. The parameter
$\sigma_V$ is referred to as the volatility of volatility, and
it measures the variance responsiveness to diffusive volatility
shocks. In the absence of jumps, the parameter $\mu$ measures the
expected log-return.

SVCJ is a rich model since it encompasses the SV and SVJ approaches.  If we set $Z_t^v=0$ in (\ref{jumpdist}), then
jumps are only present in prices, we obtain the SVJ model of \cite{bates1996jumps}.
Taking $\lambda=0$ such that jumps are not present, the model reduces
to the pure SV model originally proposed by
\cite{heston1993closed}. If we set $\kappa=\theta=\sigma_V=0$ and define $Z_t^v=0$, the
model reduces to the pure jump diffusion introduced in \cite{merton1976option}.

\subsection{Estimation: Markov Chain Monte Carlo (MCMC)}
There are plenty of different methods to estimate the diffusion process to real {data}. The generality of simulation-based methods offers obvious advantages over the method of simulated moments of \cite{duffie1993simulated}, the indirect inference methods of \cite{gourieroux1993indirect} and the efficient method of moment (EMM) method of \cite{GallantT1996}.  For example, \cite{JacquierPolsonRossi1994} %who propose a method for estimating discrete-time SV models from returns data. Their works 
show that MCMC is particularly well suited to deal with SV models. %\cite{JacquierPolsonRossi2004} extend this idea into multivariate models. 
\cite{eraker2003impact} and 
\cite{eraker2004stock} %have also developed the MCMC-based estimation of jump-diffusion models using equity returns data. \cite{eraker2003impact} 
identify several advantages of using the MCMC approach over other estimation models because MCMC methods are computationally efficient and the estimating is more flexible when using simulations. The MCMC method also provides more accurate estimates of latent volatility, jump sizes, jump times, etc. A general discussion and review of the MCMC estimation of continuous-time models can be found in \cite{johannes2009mcmc}.

For the reasons discussed {above}, we estimate the SVCJ model using the MCMC method. Doing this allows for a wide class of numerical fitting procedures that can be steered by a variation of the priors. Given that there are no BTC options yet, the MCMC method is more flexible in estimating the stochastic variance jumps and thus able to reflect
the market price of risk (\cite{franke2019statistics}. The {estimation} is based on the following Euler discretization:
\begin{eqnarray}\label{E10}
\mathrm{}Y_{t} & = & \mu + \sqrt{V_{t-1}}\varepsilon_t^y+ Z_t^yJ_t\\
V_t & = &\alpha+\beta V_{t-1}+\sigma_V \sqrt{V_{t-1}}\varepsilon_t^v+Z_t ^vJ_t,
\end{eqnarray}
where $Y_{t+1}=\log(S_{t+1}/S_t)$ is the log return, $\alpha = \kappa \theta$, $\beta = 1- \kappa$ and $\varepsilon_t^y $, $\varepsilon_t^v$ are the $\operatorname{N}(0,1)$ variables with correlation $\rho$. $J_t$ is a Bernoulli random variable with $\operatorname{p}(J_t=1)=\lambda$ and the jump sizes  $Z_t^y$ and $Z_t^v$ are distributed as specified in (\ref{jumpdist}). The daily data sample from 01/08/2014 to 29/09/2017 is used to estimate the model. All returns are in decimal form.

Let us present a brief description on how to estimate the SVCJ model with MCMC (see also \cite{johannes2009mcmc},  \cite{tsay2005analysis} and \cite{asgharian2006jump} for more details). 
Define the parameter vector as
$ \Theta= \{\mu,  \mu_y , \sigma_y, \lambda, \alpha,\beta, \sigma_v, \rho, \rho_j, \mu_v  \} $ and  $X_t =\{V_t, Z_t^y, Z_t^v, J_t\}$ as the latent variance, jump sizes and jump. Recall that $Y_t $ is the log-returns.

The MCMC method treats all components of $\Theta$ and $X \defeq \{X_t\}_{t=1,..,T}$ as random variables. The fundamental quantity is the joint pdf $p(\Theta,X|Y )$ of parameters and latent variables conditioned
on data using the Bayes formula:
\begin{eqnarray}
\operatorname{p}(\Theta,X|Y ) = \operatorname{p}(Y |\Theta,X)\operatorname{p}(X|\Theta)\operatorname{p}(\Theta).
\end{eqnarray}
The Bayes formula can be decomposed into three factors: $\operatorname{p}(Y |\Theta,X)$, the likelihood
of the data, $\operatorname{p}(X|\Theta)$ the prior of the latent variables conditioned on the parameters and $\operatorname{p}(\Theta)$ the prior of the parameters. The prior distribution $\operatorname{p}(\Theta) $ has
to be specified beforehand and is part of the model specification. In comfortable settings, the posterior variation of the parameters, given the data, is robust with respect to the prior. %This is a touchy point that is re-discussed when we display our empirical results.  

The posterior is typically not available in the closed form,
and therefore simulation is used to obtain random draws from it. This is done by generating a sequence
of draws, $ \{\Theta^{(i)}, X_t^{(i)}\}_{i=1}^N$ which form a Markov chain whose equilibrium distribution equals the posterior distribution. The point estimates of parameters and latent variables are then taken
from their sample means.

We use the same priors specified in \cite{asgharian2011risk}, who estimate a large group of international equity market returns with jump-diffusion models using the MCMC method, i.e.,
$\mu \sim \operatorname{N}(0, 25), (\alpha, \beta)\sim \operatorname{N}(0_{2\times1}, \mathbf{I}_{2\times2}),   
\sigma_2^V \sim \operatorname{IG}(2.5, 0.1),\mu_y \sim \operatorname{N}(0, 100),
\sigma_2^y \sim \operatorname{IG}(10, 40),
 \rho \sim \operatorname{U}(-1, 1),
\rho_j \sim \operatorname{N}(0, 0.5), \mu_V \sim \operatorname{IG}(10, 20)$ (Inverse Gaussian) and $\lambda \sim \operatorname{Be}(2, 40)$ (Beta Distribution). 
The full posterior distributions of the parameters and the latent-state variables can be found in \cite{asgharian2011risk} and \cite{asgharian2006jump}.
We have varied the variance of the priors and found stable outcomes, i.e., the reported mean of the posterior that is taken as an estimate of $\Theta$ is quite robust relative to changes in variance of the prior distributions.  The posterior for all parameters except $\sigma_V$ and $\rho$ are all conjugate (meaning that the posterior distribution
is of the same type of distribution as the prior but with different parameters).
The posterior for
$J_t$ is a Bernoulli distribution. The jump sizes $Z_t^y$ and $Z_t^v$ follow a posterior normal distribution
and a truncated normal distribution, respectively. Hence, it is straightforward to obtain draws
for the joint distribution of $J_t$,  $Z_t^y$ and  $Z_t^v$. However, the posteriors for $\rho$, $\sigma_2^V$
and $V_t$ are nonstandard distributions
and must be sampled using the Metropolis-Hastings algorithm. We use the random-walk
method for $\rho$ and $V_t$ , and independence sampling for $\sigma_V^2$. For the estimation of posterior moments, we perform 5000 iteratations, and in order to reduce the impact of the starting values, we allow for a burn-in for the first 1000 simulations. 

The SVCJ model is known for being able to disentangle returns related to sudden unexpected jumps from large diffusive returns caused by periods of high volatility. For the BTC situation that we consider here, we are particularly interested in linking the latent historical jump times to news and known interventions. The estimates $\hat{J_t} \defeq  (1/N) \sum_{i=1}^N J_t^i$ (where $N$ is the total number of iterations and $i$ refers to each draw) indicate the posterior probability that there is a jump at time $t$. Unlike the "true" vector of jump times, it will not be a vector of ones and zero.  Following \cite{johannes1999mcmc}, we assert that a jump has occured on a specific date $t$ if the estimated jump probability is sufficiently large, that is, greater than an approporiately chosen threshold value: 
\begin{align}
\tilde{J}_t=1\{\hat{J_t}>\zeta\}, \;\;\;\; t=1,2,...,T
\end{align} 
In our empirical study, we choose $\zeta$ so that the number of inferred jump times divided by the number of observations is approximately equal to the estimate of $\lambda$. 

%The SVCJ model is estimated with the daily BTC prices from 31/07/2014 to 29/09/2017. We first calculate returns ($R_t$)  as the log difference between BTC prices, i.e., 
%\begin{equation*}
   % $R_t=ln(P_t)-ln(P_{t-1})$, where $P_t$ is the BTC price at time $t$.
%\end{equation*}
We first estimate the BTC returns by taking the log first differences of prices, then use returns to estimate the SVCJ model. The parameter estimates (mean and variance of the posterior) of the SVCJ, SVJ and SV models for BTC are presented in Table \ref{tab:parameterbit}.  The estimate of $\mu$ is positive. The correlation between returns and volatility $\rho$ is significant and positive. This is remarkable and worth noting since it is different from a negative leverage effect observed over a sequence of studies in stock markets (see, e.g., \cite{FSS1987} and \cite{Schwert1989}). The effect is named the "inverse leverage effect" and has been discovered in commodity markets (see \cite{Schwartz2009SV}). In other words, the "inverse leverage effect" (associated with a positive $\rho$) implies that increasing prices are associated with increasing volatility. 
The reason for this positive relationship between risk and returns might be due to BTC prices being different from conventional stock prices.  The digital currency price may be dominated by the "noise trader" behavior described by \cite{Kyle1985} and \cite{Delong1990}. Such investors, with no access to inside information, irrationally act on noise as if it were information that would give them an edge. This positive leverage effect has been also reported by such as \cite{Hou2013} on other highly speculative markets, e.g., the Chinese stock markets. 

Moreover, the estimates for the SVCJ model are much less extreme than for the SVJ and SVCJ models.  More precisely, the volatility of variance $\sigma_v$ is substantially reduced from 0.017 (SV) to 0.011 (SVJ) and 0.008 (SVCJ). The mean of the jump size of the volatility $\mu_v$ is 
significant and positive. The jump intensity $\lambda$ is also significant.  The jump correlation $\rho_j$ is negative but insignificant, which parallels the results of \cite{eraker2003impact} and \cite{chernov2003alternative} for stock price dynamics. This effect might be due to the fact that even with a long data history, jumps are rare events. (The evidence is stronger for the BR specifications considered in Section 4.) In summary, the SVCJ model fits the data well by an MSE that is smaller than those of the SVJ and SV models. 

\begin{table}[h!]
\centering
 \begin{threeparttable}
\caption{BTC parameters for SVCJ, SVJ and SV models } \label{tab:parameterbit}
\scriptsize{}
 \centering
  %\begin{tabular}{P{2.5cm}P{2.5cm}P{2.5cm}}
\begin{tabular}{p{0.9cm}p{2.5cm}p{0.2cm}p{2.5cm} p{0.2cm}p{2.5cm}}
\\ \hline\hline
               & $SVCJ$& & $SVJ$&& SV    \\ [0.1cm]  \hline
  $\mu$  &     0.041    & &    0.029       & &0.030   \\  
             &[0.022, 0.060]&& [0.011, 0.046] && [0.014, 0.046]\\[0.1cm]
$\mu_y$  &     -0.084    & &-0.562       & &  - \\ 
             &[-0.837, 0.670]&& [-1.280, 0.155] && -\\[0.1cm]
$\sigma_y$& 2.155    & &2.685      & &  - \\ 
             &[1.142, 3.168]&& [1.519,3.850] && -\\[0.1cm]
$\lambda$  &     0.041    & &    0.029       & &-   \\ 
             &[0.025, 0.056]&& [0.019, 0.047] && -\\[0.1cm]
$\alpha$    &    0.010    & &    0.010      & &0.009 \\ 
             &[0.008, 0.012]&& [0.006, 0.015] && [0.006, 0.012]\\[0.1cm]
$\beta$     &    -0.132    & &   -0.116     & &-0.033 \\ 
            &[-0.151	-0.114]&& [-0.137	-0.094] && [-0.052	-0.013]\\[0.1cm]
$\rho$       &   0.407    & &  0.321    & &0.169 \\  
            &[0.232,	0.583]&& [0.225,	0.417] && [0.066,	0.271]\\[0.1cm]
$\sigma_v$  &   0.008    & & 0.011    & &0.017 \\  
            &[0.007	0.010]&& [0.007	0.014] && [0.014	0.021]\\[0.1cm]
$\rho_j$ &   -0.573   & &  -    & &- \\  
            &[-1.832, 0.685]&& - && -\\[0.1cm]
$\mu_v$ &   0.620  & &  -    & &- \\  
            &[0.426, 0.813]&& - && -\\[0.1cm]
$MSE$   &0.735	   & &0.757  &&0.763\\
\hline\hline
\end{tabular}
\begin{tablenotes}
 \footnotesize
 \item{\textit{Notes}: This table reports posterior means and 95\% finite sample credibility intervals (in square brackets) for parameters of the SVCJ, SVJ, and SV models. All parameters are estimated using BTC daily returns calculated as the log-first difference based on the prices from 01/08/2014 to 29/09/2017. }
\end{tablenotes}
\end{threeparttable}
%\end{center}
 \end{table}

\begin{figure}
	\begin{center}
	\caption{Jumps estimated in returns and volatility from the SVCJ model}\label{fig:jump1est}
	\includegraphics[width=10cm,height=8cm]{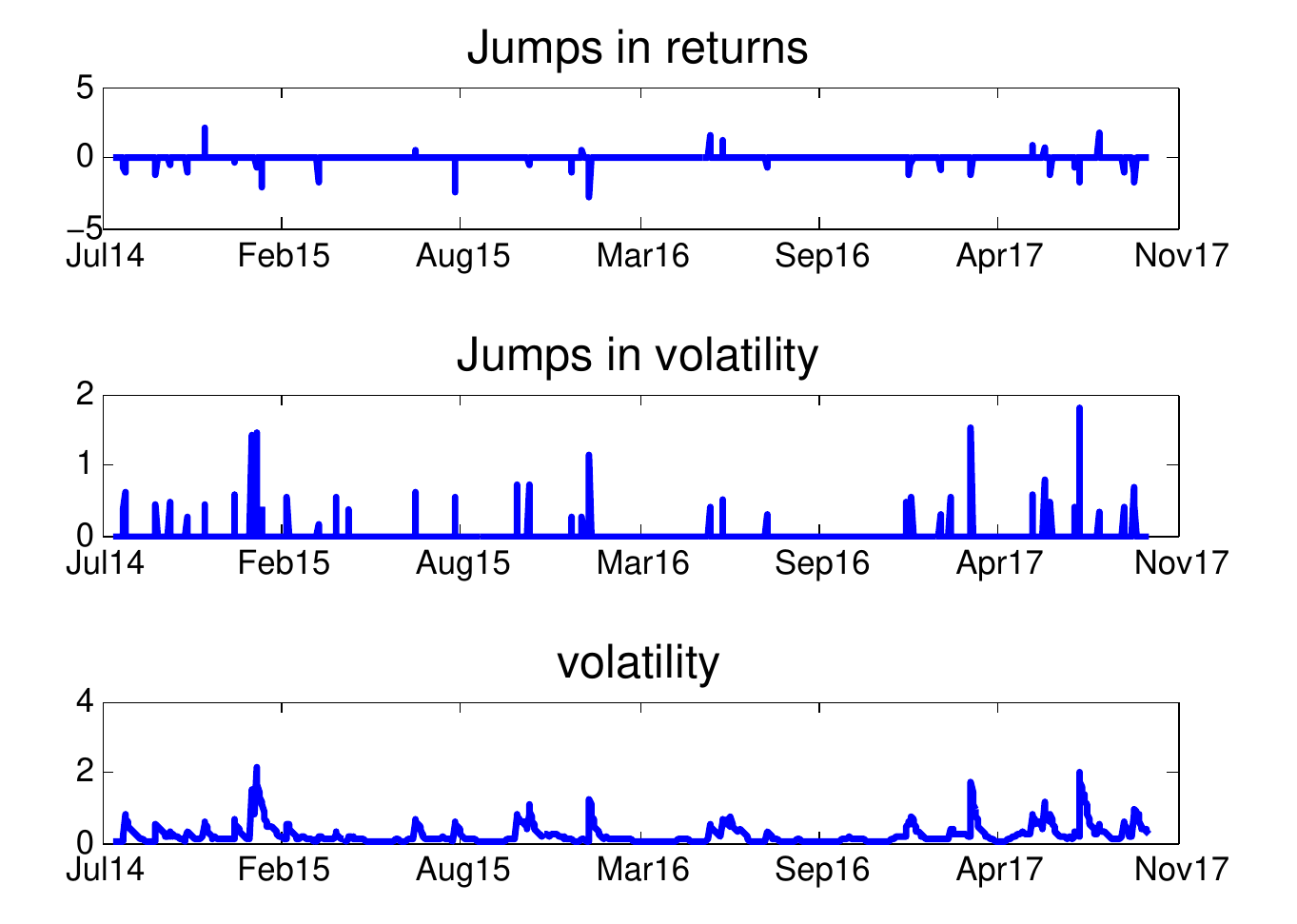}	
    \end{center}

%\begin{tablenotes}
%	\footnotesize
%	\item{\textit{Notes}: This figure graphs the estimated jumps in returns and volatility from the SVCJ model. The model is estimated using daily returns of Bitcoin from August 01, 2014 to September 29. The first sub-figure plots the jumps in returns, the  second-subfigure plots the jumps in volatility and the last sub-figure plots the estimated volatility estimated from the SVCJ model. }
%\end{tablenotes}

	\footnotesize{\textit{Notes}: This figure graphs the estimated jumps in returns and volatility from the SVCJ model. The model is estimated using BTC daily returns calculated as the log-first difference based on the prices from 01/08/2014 to 29/09/2017. The first-, second-, and third-subfigures plot jumps in returns, jumps in volatility and the estimated volatility, respectively.}	
	%	\hspace*{\fill} \raisebox{-1pt}{\includegraphics[scale=0.05]{qletlogo}\ econ\_SVCJ}
	
\end{figure}

\begin{figure}
	\begin{center}
	\caption{Estimated volatility from the SVCJ and SVJ models}\label{fig:jump2}
	\includegraphics[width=10cm,height=7cm]{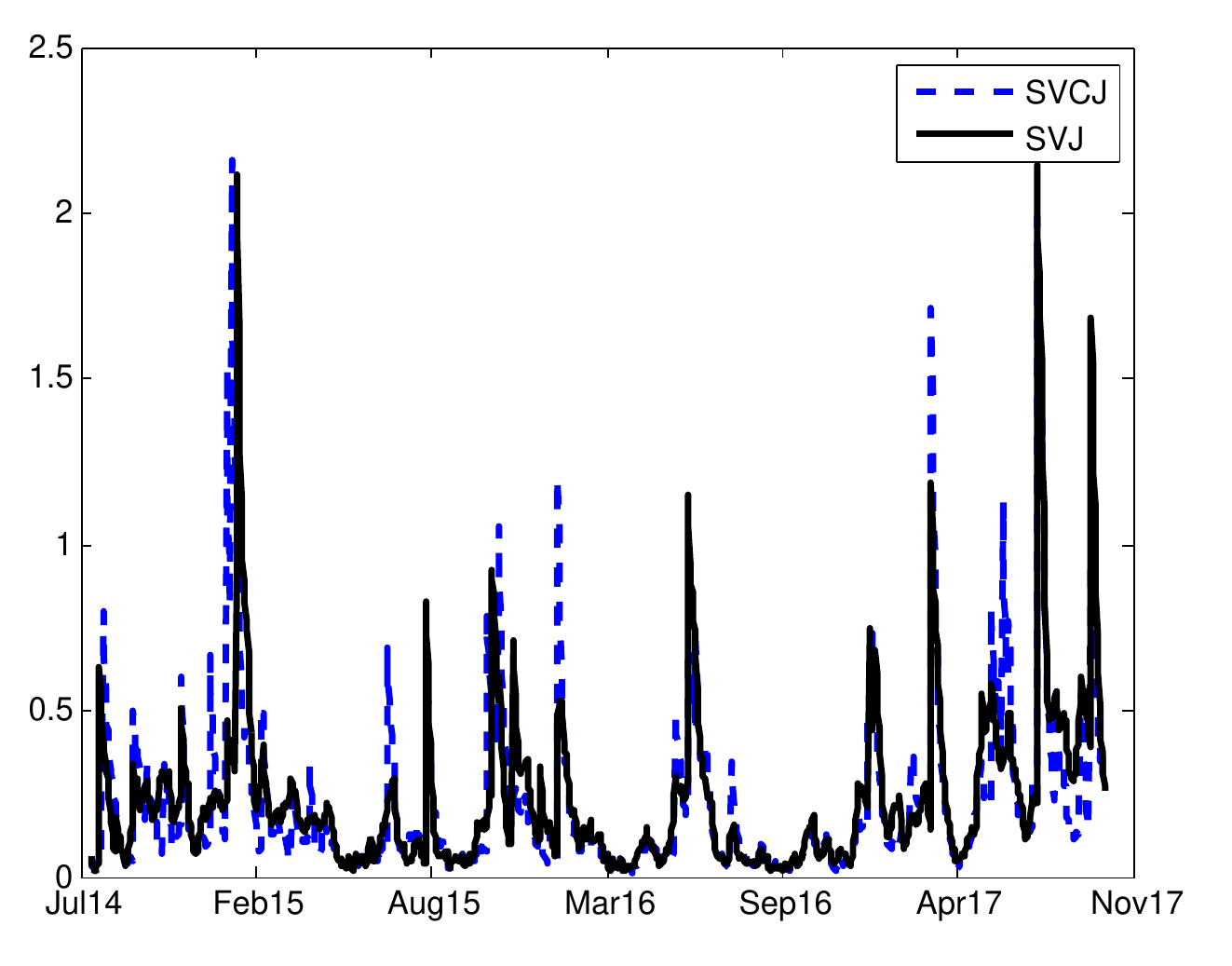}
\end{center}
\footnotesize{\textit{Notes}: This figure plots the estimated volatility from the SVCJ (dotted blue) and SVJ (solid black) models. All models are estimated using BTC daily returns calculated as the log-first difference based on the prices from 01/08/2014 to 29/09/2017.}
	%	\hspace*{\fill} \raisebox{-1pt}{\includegraphics[scale=0.05]{qletlogo}\ econ\_SVCJ}
	%\end{center}
\end{figure}

Figure \ref{fig:jump1est} shows the estimated jumps in returns (first row) and the estimated jumps in volatility (middle row) together with the estimated volatility (last row).
One sees that estimated jumps occur frequently for those of the returns and volatility. The estimated jumps size in returns and variance are different.
Figure \ref{fig:jump2} presents the in-sample fitted volatility processes for the SVCJ and SVJ models, respectively. It is not hard to see that both models lead to a similar overall pattern for the volatility process, though the SVCJ model produces sharper peaks for BTC.

A useful model diagnosis is to examine the standardized residuals obtained from the discrete model, which estimates,
\begin{align}\label{Diag}
\varepsilon_t^y=\frac{Y_t -\mu-Z_t^y J_t}{\sqrt{V_{t-1}}} 
\end{align}

\begin{figure}
	\begin{center}
		\caption{ QQ plots for the SVCJ, SVJ and SV models} \label{fig:qqbit}
	\includegraphics[width=60mm]{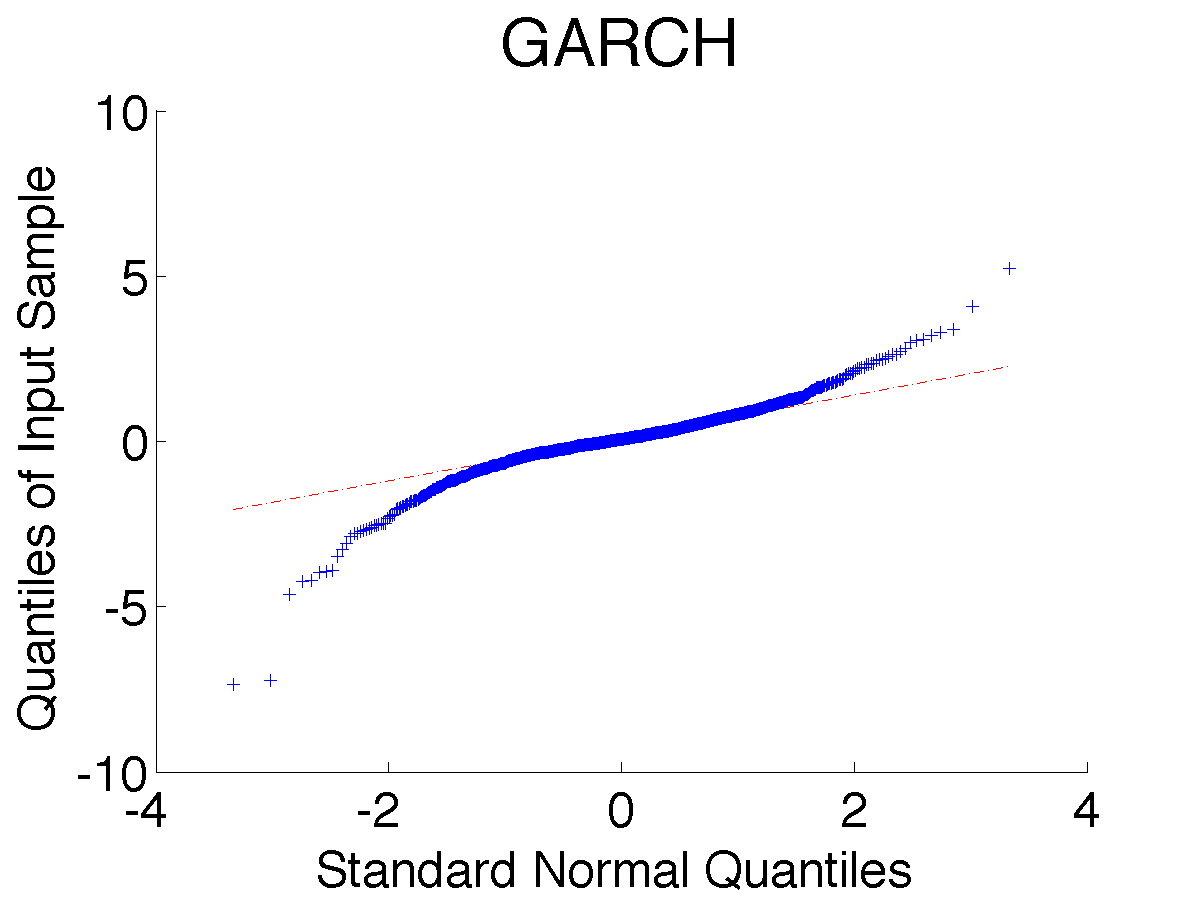}
	\includegraphics[width=60mm]{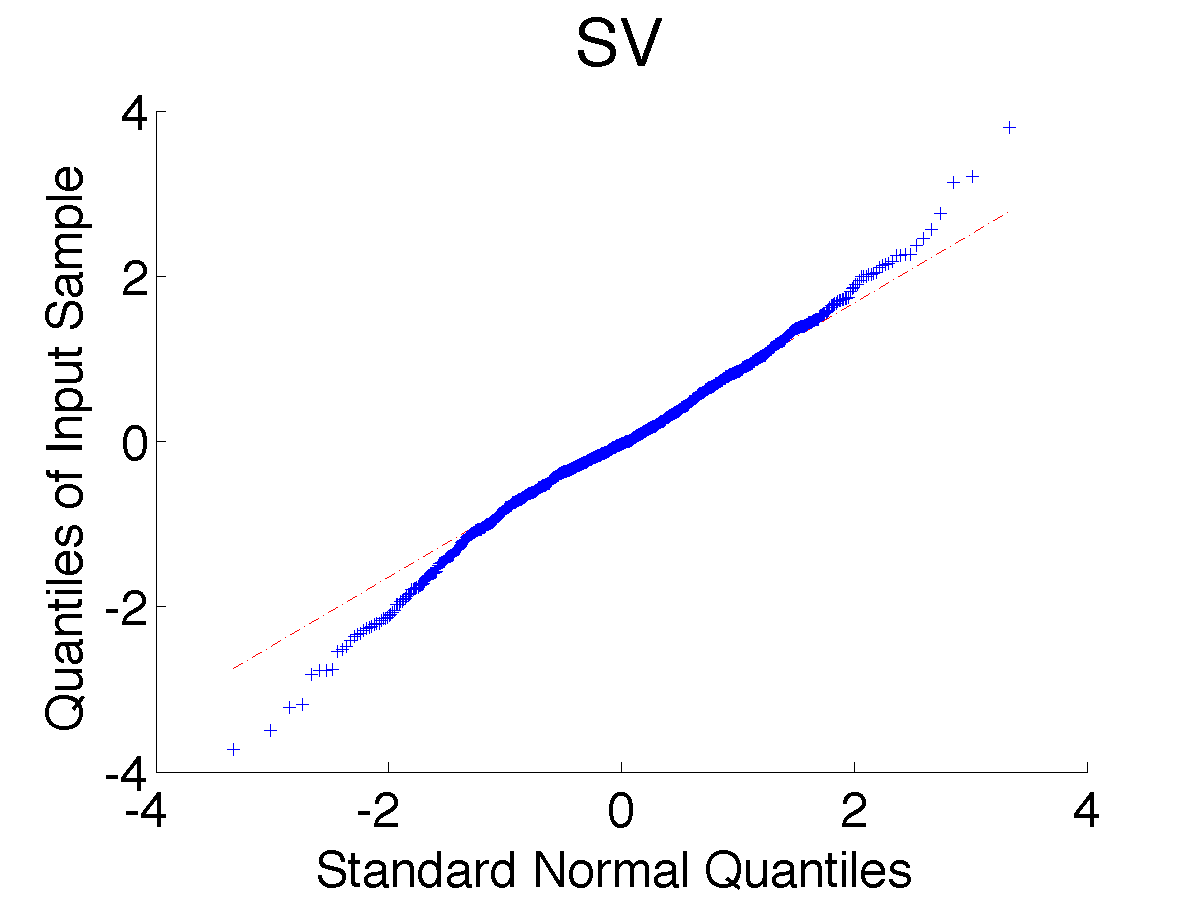} \\
	\includegraphics[width=60mm]{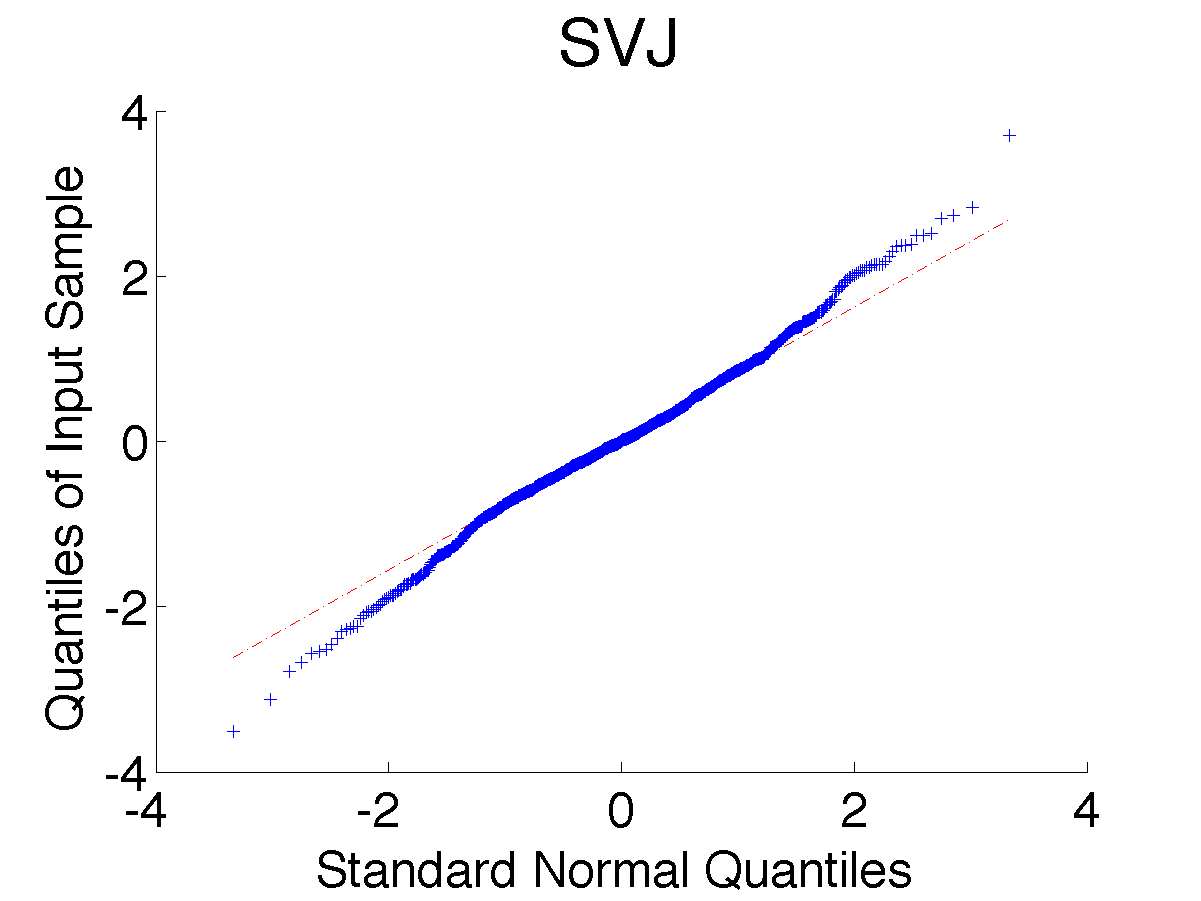}
	\includegraphics[width=60mm]{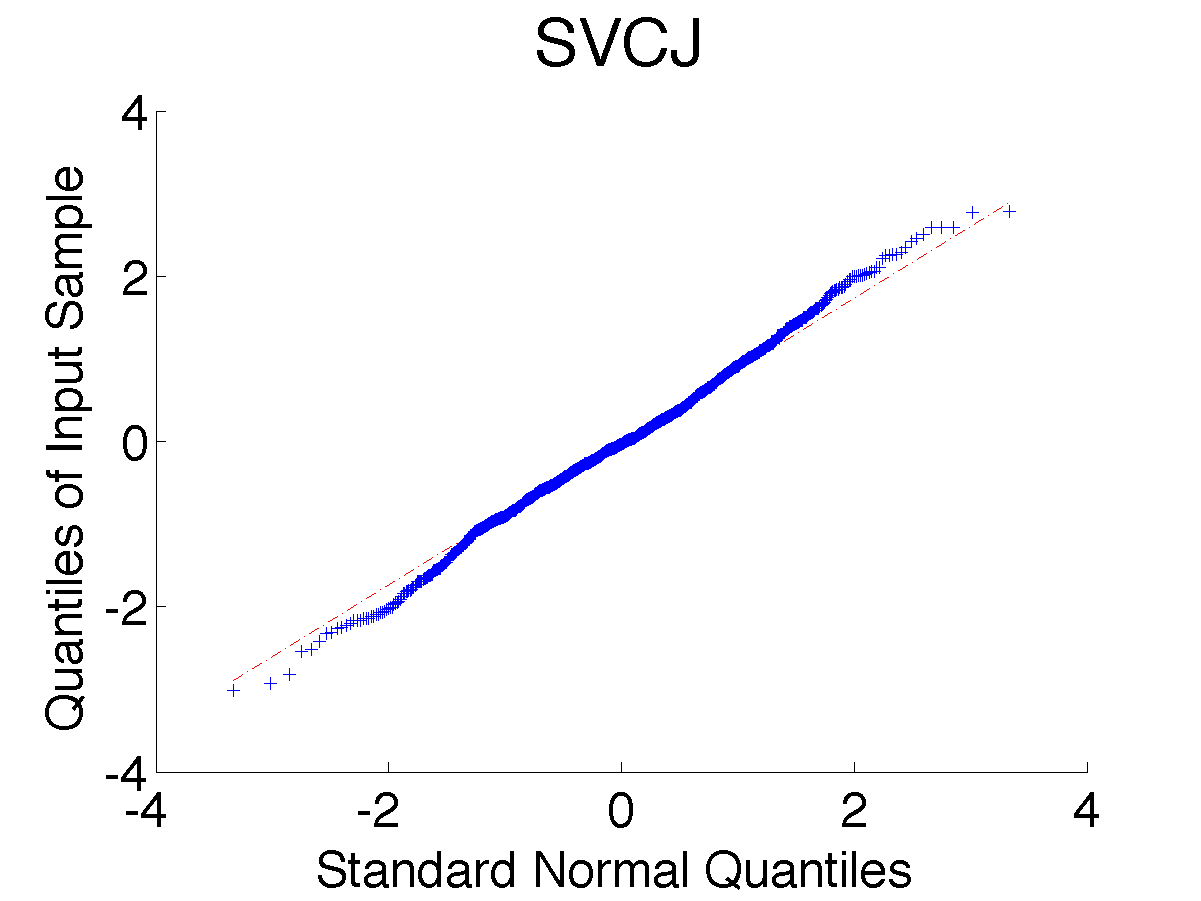}\\
	\end{center}
\footnotesize{\textit{Notes}: This figure graphs the QQ plots versus standard normal for fitted
standardized residuals from the SVCJ, SVJ and SV models using BTC daily returns calculated as the log-first difference based on the prices from 01/08/2014 to 29/09/2017. We also include the QQ plot for the GARCH model using the same sample period.}
   \end{figure}

The normality would be violated if the jumps are not perfectly estimated. However, several previous researches such as \cite{LarssonNossmanJump2011}, \cite{asgharian2006jump} and \cite{asgharian2011risk} have estimated the SVCJ model with the MCMC in the equity market and use the normal plot as a diagnostic tool to visualize the model performance. We follow these literature calculating these standardized residuals based on the estimated parameters, then show   %they should, according to Equation (\ref{Diag}), be approximately normally distributed.
 the QQ plots of the standardized residuals from the fitting of different models in Figure \ref{fig:qqbit}. From these diagnostics, it is evident that the GARCH and even the SV models are misspecified. For the SVJ and SVCJ models, the QQ plot diagnostics are substantially improved. However, it is apparent that the SVCJ model is the preferred choice which is consistent with the MSE reported in Table \ref{tab:parameterbit}.

%Finally Figure \ref{fig:jump1} graphs the 2.5th, 25th, 75th and 97.5th percentiles of the $5000$ simulated prices paths of Bitcoin for each horizon up to $30$ days based on the parameters reported in Table 4 for the SVCJ, SVJ, and SV models. The blue (red) colour line shows the $2.5th$ ($25th$) and $97.5th$ ($75th$) interval of the simulated BTC prices. The blue (red) line with a sign of $\circ$, $\diamond$, $\ast$ shows the $2.5th$ ($25th$) and $97.5th$ ($75th$) forecast intervals for the SVCJ, SVJ, and SV models, respectively. 

%Visual inspection of Figure \ref{fig:jump1} suggests that among the three models considered, the confidence intervals of BTC prices generated by the SVCJ model are narrower, particularly at the 2.5\% and 97.5\% levels. In other words, the SVCJ model predicts a narrower confidence band of extreme BTC prices than the SVJ and SV models. This implies that the SVJ and SV models produce larger upper tails, i.e., these models overestimate the average BTC price compared to the prices predicted by the SVCJ model.  On the other hand, the larger lower tails of the SVJ and SV models would imply that
%BTC would be underpriced compared to the SVCJ model.

%\begin{figure}
%	\begin{center}
%		\includegraphics[width=6cm,height=6cm]{Figures/jumpsplot}
%       \includegraphics[width=6cm,height=6cm]{Figures/jumpsplot_bit}
%		\caption{Jumps in returns and volatility from the SVCJ model for Crix(left panel) and Bitcoin(right panel).}\label{fig:jump1est}
%		\hspace*{\fill} \raisebox{-1pt}{\includegraphics[scale=0.05]{qletlogo}\ econ\_SVCJ}
%	\end{center}
%\end{figure}

% \begin{figure}
% 	\begin{center}
% 	\includegraphics[width=60mm]{figures/qqgarch}
% 	\includegraphics[width=60mm]{figures/qqsv} \\
% 	\includegraphics[width=60mm]{figures/qqsvj}
% 	\includegraphics[width=60mm]{figures/qqsvcj}\\
% 	\caption{Normal probability plots for SVCJ, SVJ, SV models for Crix.} \label{fig:qq}
%   \end{center}
% \end{figure}

% \begin{figure}
%	\begin{center}
%	\caption{Predicted confidence intervals of simulated observations for the SVCJ, SVJ and SV models  }\label{fig:jump1}
%	\includegraphics[width=7cm,height=7cm]{figures/Response_ARIMA}\\
% 		\includegraphics[width=6cm,height=6cm]{figures/Response_Crix}
 %   \includegraphics[width=12cm,height=10cm]{figures/Response_Bit_new}
  %  \end{center}
%\footnotesize{\textit{Notes}: This figure graphs the confidence intervals of simulated BTC price paths up to $30$ days based on the parameters reported in Table 4 for the SVCJ, SVJ and SV models. The blue (red) line shows $2.5th$ ($25th$) and $97.5th$ ($75th$) percentiles of the simulated BTC prices. The blue (red) line with a sign of $\circ$, $\diamond$, $\ast$ plots the $2.5th$ ($25th$) and $97.5th$ ($75th$) percentile for the SVCJ, SVJ and SV models, respectively.}
		%\hspace*{\fill} \raisebox{-1pt}{\includegraphics[scale=0.05]{qletlogo}\ econ\_SVCJ}

%\end{figure}

\section{SV model with jumps: high frequency data}\label{sec:nonAffine}

\subsection{BR model in return-volatility co-jumps } \label{sec:BR}
Imposing a specific structure in the stochastic process as documented in Section \ref{sec:affine} may produce a specification error. Defining $S_t$ and $\sigma_t = \sqrt{V}_t$ as the price and volatility process, respectively, following the notation of BR, we therefore consider the BR affine jump-diffusion model:
\begin{equation}
\label{jumppar}
\begin{split}
d\log(S_t) & = \mu_r dt + \sigma_t\{\rho_tdW_t^1 + \sqrt{1 - \rho_t^2}dW_t^2 \} \\
& + c_{r,t}^JdJ_{r,t} + c_{r,t}^{J\!J}dJ_{r, \sigma,t},\\
d\xi(\sigma_t^2) & =\{m_0+ m_1\log(\sigma_t^2)\}dt + \Lambda dW_t^1 + c_{\sigma, t}^JdJ_{\sigma,t} + c_{\sigma, t}^{J\!J}dJ_{r, \sigma,t},\\
\rho_t &= \mbox{max}\{\min(\rho_0+\rho_1\sigma_t,1),-1\},
\end{split}
\end{equation}
where $\xi(\cdot)$ is an increasingly monotonic function (we will choose it as $\log(\cdot)$ in the following discussions), $W = \{W^1, W^2\}$ is a bivariate standard Brownian motion vector and $J = \{J_{r,t}, J_{\sigma,t}, J_{r, \sigma,t} \}$ is a vector of mutually independent Poisson processes with constant intensities, which are denoted as $\lambda_r$, $\lambda_{\sigma}$ and $\lambda_{r, \sigma}$, respectively. 
Thus we allow for common and independent jumps in the system. The Poisson processes are also assumed to be independent from the Brownian motion. 

%In order to guarantee the existence and uniqueness of a solution to the system, it is assumed by BR that functions $\mu(\cdot)$, $m(\cdot)$, $\Lambda(\cdot)$, $\lambda_r(\cdot)$, $\lambda_{\sigma}(\cdot)$, $\lambda_{r, \sigma}(\cdot)$ and $\rho(\cdot)$ meet certain mild regularity conditions.
% Furthermore, the recurrence of the variance also hold in the model.

%The BR model is estimated through the GMM method via estimated infinitesimal cross-moments. 
The BR model is estimated through a GMM-like procedure based on infinitesimal cross-moments dubbed by the authors NIMM or Nonparametric Infinitesimal Method of Moments.
We assume the distribution of the jumps to be normal, i.e. $(c_{r,t}^{J}, c_{\sigma,t}^{J}) \sim \N(\mu^J, \Sigma^J)$ and $(c_{r,t}^{J\!J}, c_{\sigma,t}^{J\!J}) \sim \N(\mu^{J\!J}, \Sigma^{J\!J})$, with
\begin{equation}
\begin{split}
&\mu^{J} = \begin{bmatrix}\mu_{J,r}\\\mu_{J,\sigma}\end{bmatrix}, \quad\mu^{J\!J} = \begin{bmatrix}\mu_{J\!J,r,0}+ \mu_{J\!J,r,1}\sigma_t\\\mu_{J\!J,\sigma}\end{bmatrix},\\
&\Sigma^{J} = \begin{bmatrix}\sigma^2_{J,r} & 0\\ 0 & \sigma^2_{J,\sigma}\end{bmatrix}, \quad \Sigma^{J\!J} = \begin{bmatrix}(\sigma_{J\!J,r,0}+\sigma_{J\!J,r,1}\sigma_t^{\sigma_{JJ,r,2}})^2 & \rho_J(\sigma_{J\!J,r,0}+\sigma_{J\!J,r,1}\sigma_t^{\sigma_{JJ,r,2}})\sigma_{J\!J,\sigma}\\ \rho_J(\sigma_{J\!J,r,0}+\sigma_{J\!J,r,1}\sigma_t^{\sigma_{JJ,r,2}})\sigma_{J\!J,\sigma} & \sigma^2_{J\!J,\sigma}\end{bmatrix}.
\end{split}
\end{equation}
For any $p_1 \geq p_2 \geq 0$, the generic infinitesimal cross-moment of order $p_1$ and $p_2$ is defined as:
\begin{equation}
\theta_{p_1, p_2}(\sigma) = \lim_{\Delta \to 0} \frac{1}{\Delta} \E \{[\log(S_{t + \!\Delta}) - \log(
S_{t})]^{p_1}[\log(\sigma^2_{t + \!\Delta}) - \log(\sigma^2_{t})]^{p_2}|\sigma_t = \sigma\}.
\end{equation}
In particular $\theta_{p_1, 0}$ helps to identify features of the price process, and $\theta_{0, p_2}$ helps to identify those of the variance process, while the genuine cross-moments with $p_1 \geq p_2 \geq 1$ are required to identify the common parameter shared by the two processes $\rho_0,\rho_1$, $\lambda_{r, \sigma}$ and $\rho_J$.

%To conduct the GMM estimation in BR, 
To conduct the NIMM estimation in BR, we first need to estimate  the cross-moments that are in theory functions of parameter of interest. The cross-moments are estimated via a nonparametric kernel method. In particular, denote the day index as $t = 1, ..., T$ and the equispaced time index as $i = 1, ..., N$ within each day. Denote $r_{t,i,k}$ as the high-frequency log returns for day $t$, knot $i$ and minute $k$. We define the closing logarithmic prices as $\log(p_{t,i})$ and logarithmic spot variance estimates as
\begin{equation} \label{spot}
\hat{\sigma}_{t,i}^2 = \frac{T}{T - 1 - n_j}  \zeta_1^{-2} \sum_{k=2}^{T}|r_{t,i,k}||r_{t,i,k-1}| \mathbbm{1}_{\{|r_{t,i,k}| \leq \theta_{t,i,k} \}}\mathbbm{1}_{\{|r_{t,i,k-1}| \leq \theta_{t,i,k-1} \}},
\end{equation}
where $\zeta_1 \approx0.7979$, $\theta_{t,i,k}$ is a suitable threshold, and $n_j$ is the number of returns whose absolute value is greater than $\theta_{t,i,k}$. Then the generic cross-moment estimator $\hat{\theta}_{p_1, p_2}(\sigma)$ is defined as
\begin{equation}
\hat{\theta}_{p_1, p_2}(\sigma) = \frac{\sum_{t=1}^{T-1}\sum_{i=1}^NK(\frac{\hat{\sigma}_{t,i} - \sigma}{h})\{\log(S_{t + 1, i}) - \log(S_{t,i})\}^{p_1}  \{\log(\hat{\sigma}^2_{t + 1, i}) - \log(\hat{\sigma}^2_{t, i})\}^{p_2}}{\Delta\sum_{t=1}^T\sum_{i=1}^N K(\frac{\hat{\sigma}_{t,i} - \sigma}{h})}
\end{equation}
where $K(\cdot)$ is a kernel function and $h$ is the bandwidth. Finally, with the estimated cross-moments, one can estimate the parameters of interest via the NIMM method, see the details as in \cite{bandi2016price} for the parametric estimation.
%The procedure is actually two-step in the sense that spot variance is estimated first and then kernel estimates of the cross-moments are used to identify the system's dynamics with the GMM method.

\subsection{Correspondence between SVCJ and BR model}
In this section, we fit the BR model using high-frequency data and discuss the comparison with the {estimation} of the SVCJ model.
We collect high-frequency BTC prices from Bloomberg. The price data range is from $31/07/2014$ to $29/07/2017$, and we collect raw data at a frequency of $60$ seconds 24 hours a day. Following Section \ref{sec:BR}, we aggregate the logarithm returns of Bitcoin over a $60$-minute time range, namely $r_{t,i,k} = \log S_{t,i,k}- \log S_{t,i,k-1}$, with $k= 1, \cdots, 60$. In addition, we also obtain the spot variance estimates for each day $t$ and each knot $i$ by applying the jump robust threshold bipower variation estimator as in Equation (\ref{spot}).
%\begin{equation}
%\hat{\sigma}^2_{t,i} = \frac{60}{59-n_j}\varsigma^{-2}\sum^{65}_{k=2}|r_{t,i,k}r_{t,i,k-1}| \1\{|r_{t,i,k}|\leq \theta_{t,i,k}\}\1\{|r_{t,i,k-1}|\leq \theta_{t,i,k-1}\},
%\end{equation}
%where $\varsigma$ is around $0.7979$, and $\theta_{t,i,k}$ is a suitable threshold, and $n_j$ is the number of returns whose absolute value is greater than the threshold.

\begin{figure}[h]
\begin{center}
\caption{The averaged daily spot volatility and the daily spot volatility estimates.}
\centering
\label{figure:var}
\includegraphics[width=1\textwidth]{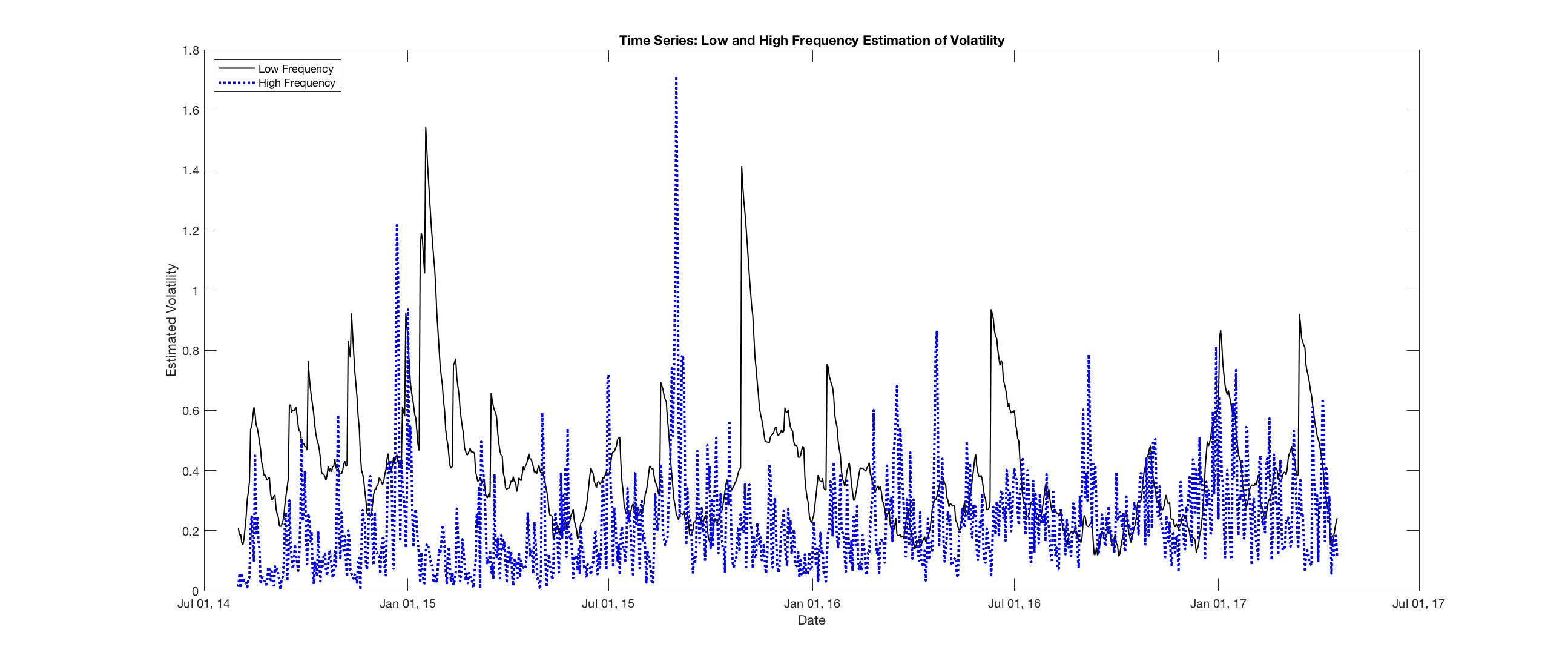}
\end{center}
\footnotesize{Notes: This figure plots the averaged daily spot volatility from the high-frequency data (dotted line in blue) and the daily spot volatility estimates (solid line in black)  }
\end{figure}

To compare the data of the high-frequency aggregated volatility and the daily Bitcoin volatility, we plot the averaged daily spot volatility from the high-frequency data and the daily spot volatility estimates from the SVCJ model together as in Figure \ref{figure:var}. We observe that the two sequences sometimes peak at different time points despite that the general pattern agrees. 

In Table \ref{tabparameterbandi}, we show the full model estimation results. 
The drift parameter $\mu_r$ is estimated to be small and insignificant.  The linear mean reversions, which can be seen as $m_0$ and $m_1$, are both negative. However they are both insignificant. The volatility of volatility $\Lambda$ is estimated to be very significant with a value of $0.6766$. The averaged number of independent jumps in volatility is estimated at an annual rate of $0.0519* 252$, which is around $13$. The estimated number of co-jumps is around $0.0584* 252 \approx 17$. The mean of the independent variance jumps is significant at a level of $-0.2783$. $\mu_{JJ,r,0}$ is small (-0.0187) and negative, and $\mu_{JJ,r,1}$ is $0.1265$. Both parameters are insignificant at the $95\%$ level of confidence. We do not see an obvious tendency for the jumps to be downward, as observed in   \cite{bandi2016price}.

We find that the leverage $\rho_0$ is estimated to be negative, i.e., $-0.1485$, though insignificant. The leverage would increase with an increasing volatility level as $\rho_1$ is estimated to be significant and with a value of  $0.9292$. The standard deviation of the jumps in return $\sigma_{J,r}$ is estimated to be significant with the value of $0.6890 $. When fitting a nonlinear structure to the standard deviation of the common price jumps, the parameters $\sigma_{JJ,r,1}$ and $\sigma_{JJ,r,2}$  are both significant. The standard deviation of jumps in volatility $\sigma_{J,\sigma}$ is estimated to be $0.8619$ with significance. The standard deviation of the common volatility jump $\sigma_{JJ,\sigma}$ is estimated to be insignificant. Notably, the correlation of jumps $\rho_J$ is estimated to be negative and significant with a value of $-0.5257$, which is in line with BR. This negative and significant co-jump size correlation is discovered by \cite{duffie2000transform}, who conclude that the price and the volatility jump sizes are "nearly perfectly anti-correlated". \cite{eraker2004stock} finds a statistically significant correlation between the jump sizes only when employing option data in addition to stock returns data. \cite{bandi2016price} also report a "nearly perfect anti-correlation" of -1.

\begin{table}[h!]
\centering
 \begin{threeparttable}
 \centering
\caption{BR parametric estimates and their $95\%$ confidence intervals. The parametric model is specified as in Equation \ref{jumppar}-12. The first column specifies that $J_{r,\sigma} = 0$ (no co-jumps) and the second column specifies that $J_r= J_{\sigma} = 0$ (no independent jumps).} \label{tab:parameterbitBR}
\scriptsize{}
 \centering
  \label{tabparameterbandi}
  %\begin{tabular}{P{2.5cm}P{2.5cm}P{2.5cm}}
\begin{tabular}{p{0.9cm}p{2.5cm}p{0.2cm}p{2.5cm}p{0.2cm}p{2.5cm}}
\\ \hline\hline
                        & no cojumps       && no ind. jumps        && full model       \\[0.1cm]
                        \hline
$\mu_r$                 & 0.0021            && 0.0027               && 0.0082               \\
                        & [-0.1939, 0.1981] && [-0.1933, 0.1987]    && [-0.0444, 0.0608]    \\[0.1cm]
$\rho_0$                & 0.0044            && -0.0148              && -0.1485              \\
                        & [-0.1150, 0.1237] && [-0.1401, 0.1105]    && [-0.4851, 0.1882]    \\[0.1cm]
$\rho_1$                & -0.3744           && -0.2237              && 0.9292               \\
                        & [-0.8513, 0.1025] && [-0.7088, 0.2614]    && [0.5884, 1.2699]     \\[0.1cm]
$m_0$                   & -0.0500           && -0.0500              && -0.0495              \\
                        & [-0.1275, 0.0275] && [-0.1275, 0.0275]    && [-0.1475, 0.0485]    \\[0.1cm]
$m_1$                   & -0.0168           && -0.0125              && -0.0600              \\
                        & [-0.2128, 0.1792] && [-0.2085, 0.1835]    && [-0.2560, 0.1360]    \\[0.1cm]
$\Lambda$               & 0.7634            && 0.7853               && 0.6766               \\
                        & [0.5674, 0.9594]  && [0.5893, 0.9813]     && [0.6570, 0.6963]     \\[0.1cm]
$\mu_{J,r}$             & 0.1577            && 0                    && 2.5486               \\
                        & [0.0372, 0.2782]  && -                    && [2.3526, 2.7446]     \\[0.1cm]
$\mu_{JJ,r,0}$          & 0                 && -0.0804              && -0.0187              \\
                        & -                 && [-0.5383, 0.3774]    && [-0.1085, 0.0711]    \\[0.1cm]
$\mu_{JJ,r,1}$          & 0                 && 0.0192               && 0.1265               \\
                        & -                 && [-0.6850, 0.7234]    && [-0.4183, 0.6713]    \\[0.1cm]
$\sigma_{J,r}$          & 0.6801            && 0                    && 0.6890               \\
                        & [0.5453, 0.8148]  && -                    && [0.4930, 0.8850]     \\[0.1cm]
$\sigma_{JJ,r,0}$       & 0                 && 0.0864               && 0.0043               \\
                        & -                 && [-0.3242, 0.4971]    && [-0.5459, 0.5544]    \\[0.1cm]
$\sigma_{JJ,r,1}$       & 0                 && 1.8713               && 1.2159               \\
                        & -                 && [1.8436, 1.8991]     && [1.0199, 1.4119]     \\[0.1cm]
$\sigma_{JJ,r,2}$       & 0                 && 2.6521               && 3.9590               \\
                        & -                 && [2.5377, 2.7664]     && [3.7630, 4.1550]     \\[0.1cm]
$\mu_{J,\sigma}$        & -0.5000           && 0                    && -0.2783              \\
                        & [-0.5364,-0.4636] && -                    && [-0.4992, -0.0574]   \\[0.1cm]
$\mu_{JJ,\sigma}$       & 0                 && -1.9181              && -0.4927              \\
                        & -                 && [-2.0805,-1.7557]    && [-0.6429, -0.3425]   \\[0.1cm]
$\sigma_{J,\sigma}$     & 0.7945            && 0                    && 0.8619               \\
                        & [0.7379, 0.8511]  && -                    && [0.7237, 1.0001]     \\[0.1cm]
$\sigma_{JJ,\sigma}$    & 0                 && 1.0705               && 0.0717               \\
                        & -                 && [0.8716, 1.2693]     && [-0.0767, 0.2202]    \\[0.1cm]
$\rho_J$                & 0                 && -1.0000              && -0.5257              \\
                        & -                 && [-1.4648, -0.5351]   && [-0.7217, -0.3297]   \\[0.1cm]
$\lambda_r$             & 0.0002            && 0                    && 0.0000               \\
                        & [-0.1958, 0.1962] && -                    && [-0.1960, 0.1960]    \\[0.1cm]
$\lambda_\sigma$        & 0.0700            && 0                    && 0.0519               \\
                        & [0.0504, 0.0896]  && -                    && [0.0323, 0.0715]     \\[0.1cm]
$\lambda_{r,\sigma}$    & 0                 && 0.0060               && 0.0584             \\
                        & -                 && [-0.0136, 0.0256]    && [0.0564, 0.0603]     \\[0.1cm]
\hline\hline
\end{tabular}
\begin{tablenotes}
\item \textit{Notes}: This table reports the parameter estimates of the model specified in Equation 11-12 using the intra-daily BTC returns.  For each parameter, we report the estimate and the corresponding $95\%$ finite sample credibility intervals in parentheses. The full model is shown in the forth column, and the second and third columns report the same model with the restriction of no co-jumps and no independent jumps, respectively. 
\end{tablenotes}
\end{threeparttable}
%\end{center}
\end{table}

% % \begin{table}[]
% \caption{Correspondences between estimated parameters}
% \label{comp}
% \centering
% \begin{tabular}{r|r|r|r}
% \hline\hline
% {Hou et.al., (2019)}                                                                              & &  \cite{bandi2016price} &  \\ \hline
% $\lambda$                                                                                        & $0.041^{**}$     & $\lambda_{r, \sigma}$ & $0.0583^{**}$         \\ \hline
% $\sigma_v$                                                                                       & $0.008^{**}$     & $\Lambda$             & $0.6766^{**}$   \\ \hline
% $\mu$                                                                                            & $0.041^{**}$     & $\mu_r$                 & $0.0082$   \\ \hline
% $\rho_j$              &        $-0.573$     & $\rho_J$                 & $-0.5257^{**}$  \\ \hline\hline
% \end{tabular}
% \end{table}

% \ref{figurenon} shows fitted results of the parametric functions. We find that there are small changes of the parameter values in the region of low volatility level, for example  $\mu_{JJ,r},\mu_{JJ,s}$, $\sigma_{JJ,r},\sigma_{JJ,s}$ and $\lambda_{JJ}$, then it stays constant for the majority of the parameter region. Therefore we can comfortably fit a parametric model similar to \cite{bandi2016price}.
%\begin{figure}[h]
%\begin{center}
%\centering
%\includegraphics[width=1.1\textwidth]{result_nonparametric.png}
%\caption{Parametric fitting results (solid line) with estimated confidence band (dotted line).}
%\label{figurenon}
%\end{center}
%\end{figure}

\section{Option pricing}\label{sec:opt}
In the previous sections, we have shown that the SVCJ and the the BR models can well describe the log returns dynamics of BTC. In this section, we discuss option pricing for BTC based on the SVCJ and BR models, respectively. 

\subsection{BTC options}
%Derivative securities such as futures and options are priced under a probability measure $Q$ commonly referred to as the ``risk neutral'' or martingale measure. Under $Q$ the spot price and volatility dynamics are given by,
%\begin{eqnarray}
%\frac{\mathrm{d}S_t }{S_t}& = &(r-\mu^* )\mathrm{d}t+\sqrt{V_t}\mathrm{d}W_t^{(S,Q)}+Z_t^y\mathrm{d}N_t^{(Q)}\\
%\mathrm{d}V_t & = &
%\kappa(\theta-V_t)\mathrm{d}t+\sigma_V\sqrt{V_t}\mathrm{d}W_t^{(V,Q)}+Z_t^v\mathrm{d}N_t^{(Q)}
%\end{eqnarray}

%where $W^{S,Q}$, $W^{V,Q}$ and $N^{S,Q}$ , $N^{V,Q}$ are the corresponding Brownian motions and Poisson process under Q. The absence of arbitrage requires that the drift of $S$ under $Q$ is given by {$(r-\mu^*) $} where $r$ is the risk free interest rate and $\mu^*=\E{_Q}\{\exp(Z_t^Y)\}-1$ is the jump compensator. Since our purpose is to explore the impact of model choice on option values we follow \cite{eraker2003impact} and set the risk premia to be 0. This is disputable, but for the lack of existence of the officially traded options a justifiable path to pricing BTC option contingent claims. 

After {fitting} the SVCJ and the BR model, {we advance} with a numerical technique called Crude Monte Carlo (CMC) to approximate the BTC option prices. Derivative securities such as futures and options are priced under a probability measure Q commonly referred to as the “risk neutral” or martingale measure. Since our purpose is to explore the impact of model choice on option prices, we follow \cite{eraker2003impact} and set the risk premia to zero. This choice can be disputed, but for the lack of existence of the officially traded options a justifiable path to pricing BTC contingent claims.
Suppose we have an option with a payoff at time of maturity $T$ as $C(T)$, and typically for call option $C(T) = (S_T -K)^{+}$.
The price of this option at time $t$ is denoted as: 
\begin{equation}\label{eq22}
\E{_Q} [\exp \{ -r(T-t) \} C(T)|\mathcal{F}_t],
\end{equation}
where $\mathcal{F}_t$ is a set that represents information up to time $t$.
We approximate the European option prices of BTC using the CMC technique. The CMC simulation is done for {20000} iterations to approximate the option price using the parameters reported in Table \ref{tab:parameterbit} for the SVCJ, SVJ, and SV models and {in Table \ref{tab:parameterbitBR}} for the BR (assuming a daily interval) model. Since no BTC option market exists yet, we do not have real market option prices for comparison. Thus, we chose July 2017 randomly as the experimental month in our option-pricing simulation analysis. Throughout our entire analysis of option pricing, the moneyness for strike $K$ and $S$ at $t$ is defined to be $K/S_t$. The pricing formula is a function of moneyness and time to maturity $\tau = (T-t) $ where $T$ is the maturity day.

\begin{figure}
	\begin{center}
		\caption{Estimated volatility of BTC for July 2017: BTC}\label{fig:vol}
        \includegraphics[width=12cm,height=8cm]{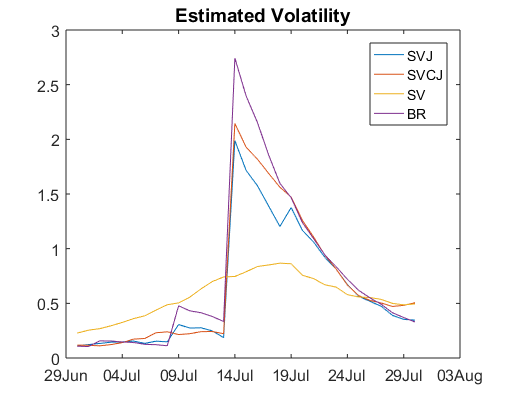}
    	\end{center}
    	
\footnotesize{\textit{Notes}: This figure plots the estimated volatility of the SVCJ, SVJ, SV and the BR models. The volatility is approximated based on the parameters reported in Table \ref{tab:parameterbit} and {Table \ref{tab:parameterbitBR}} for the month of July 2017. The x-axis notes the dates in July 2017. {The blue/red/orange/purple line plots the volatility from the SVJ/SVCJ/SV/BR models. }}
	%	\hspace*{\fill} \raisebox{-1pt}{\includegraphics[scale=0.05]{qletlogo}\ econ\_SVCJ}

\end{figure}

In Figure \ref{fig:vol}, we plot the simulated volatility of various models based on the parameters reported in Table \ref{tab:parameterbit} {(for the SVCJ, SVJ and SV models)} and  {in Table \ref{tab:parameterbitBR} (for the BR model)} for the month of July 2017. It can be seen from this figure that the approximated volatility on 15 July, 2017, had a large jump (there was a large increase observed on 15 July, 2017, in the BTC historical prices). The sudden jump is perfectly captured by the BR, SVCJ and SVJ models, while the SV model cannot characterize the volatility as well as the other three models. The BR model estimates the jump more than the SVCJ {and} the SV model, this could be attributed to the uncorrelated jumps, which is not considered by the SVCJ and the SVJ models.  Assuming a BTC spot price $S_t$ = 2250, the estimated BTC call option prices across moneyness and time to maturity on July 17, 2017, obtained using the SVCJ model\footnote{We have also calculated option prices for the SVJ, SV models. These results are available upon requests. The codes for this research can be found in {\tt www.quantlet.de}.} are presented in Table \ref{optbitsvcj}. We see that, for example, a call option on BTC with the strike  $K = 1250$ and time to maturity of 90 days would be traded at 1157.95 on 17 July, 2017. %It is obvious from Table \ref{optbitsvcj} that the BTC option prices increase with time to maturity and decrease across moneyness from ITM to OTM. %This is consistent with option prices in the real world and also observed in the equity option markets. 

%We show also the movements of approximated BTC call option prices using the parameters reported in Table 4 for July 2017 in Figure \ref{fig:price}. In this figure, the estimated price of BTC call option changes with respect to changes in moneyness (assuming 30 days time to maturity) and across different time to maturity (assuming the strike price is 2250). It can be seen from this figure that BTC call option prices decrease across moneyness from ITM to OTM and increase when time to maturity increases. 

\begin{figure}
\begin{subfigure}
  \centering
  \caption{Call option prices across moneyness and time to maturity: BTC}
    \includegraphics[width=8cm,height=8cm]{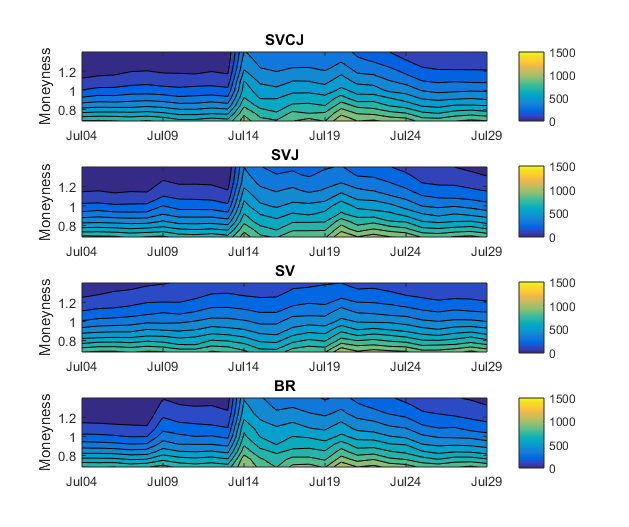}
   \includegraphics[width=8cm,height=8cm]{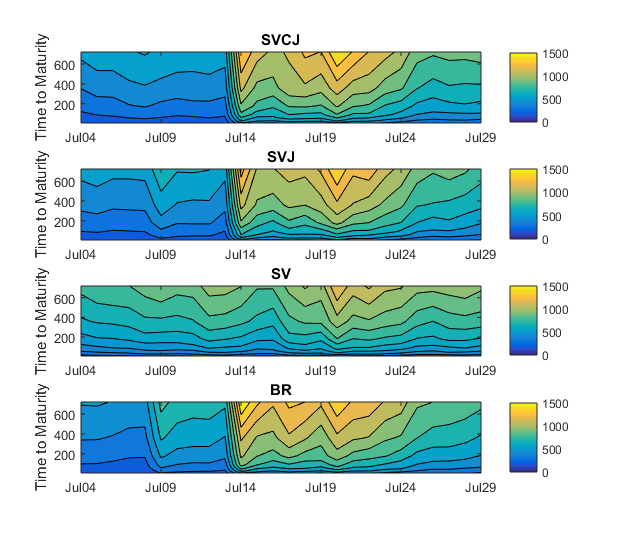}
\footnotesize{\textit{Notes}: This figure graphs the call option prices surface counter plot across different moneyness and different times to maturities for the month of July 2017, as shown in the right-hand side labels. When looking at moneyness, the time to maturity is fixed at 30 days, and when looking at the time to maturity, moneyness is ATM. The colour in the graph represents the price level; the brighter the colour, the higher the price. } 
  \label{fig:money}
\end{subfigure}%
%\begin{subfigure}
 % \centering
  %\includegraphics[width=8cm,height=6cm]{figures/TimetoMaturity}
  % \includegraphics[width=8cm,height=6cm]{figures/TimetoMaturity_b%it}
 % \caption{Call option price against time to maturity for BTC %on 201707.}
%  \label{fig:ttoMaturity}
%\end{subfigure}
%\caption{plots of....}
%\label{fig:fig}
\end{figure}

To further understand how the option price changes with respect to changes in time to maturity and moneyness for different models, we show in Figure \ref{fig:money} the one-dimensional contour plot of the option prices surface across time to maturity and moneyness estimated from the SVCJ, SVJ, SV and the BR models for the month of July 2017. When examining moneyness, the time to maturity is fixed at 30 days, and when looking at the time to maturity, moneyness is fixed at at-the-money (ATM). We can see from the contour plot that the relationship between the option price and the time to maturity or moneyness varies over time for all four models. %However, one easily seen pattern is that the approximated option price is higher when the volatility is higher, i.e., the colour of the contour plot is brighter. This is especially the case for options varying across time to maturity. %The jump in volatility on July 15, 2017 can also be observed easily in the prices coutor plots based on the SVCJ and SVJ models, there is a sharp increase in the curve on july 15, 2017. 
The BR model and the SVCJ models have more volatile {patterns} than those of the SCJ and SV models. This figure conveys a homogeneous message as we can see from Figure \ref{fig:vol} in the volatility plots. For example, for the BTC price, we see a drastic change in the contour structure on, e.g., 15 July, 2017 as the price suddenly drops from  2232.65 USD on $15/07/2017$ to 1993.26 USD. The sudden drop in price should be attributed to the big jump in volatility shown in Figure \ref{fig:vol}, and we can also observe this jump on 15 July in Figure \ref{fig:money}. 

\begin{figure}
\begin{subfigure}
  \centering
   \caption{Call option price differences between the SVCJ and SVJ models: BTC}
  \includegraphics[width=8cm,height=5.5cm]{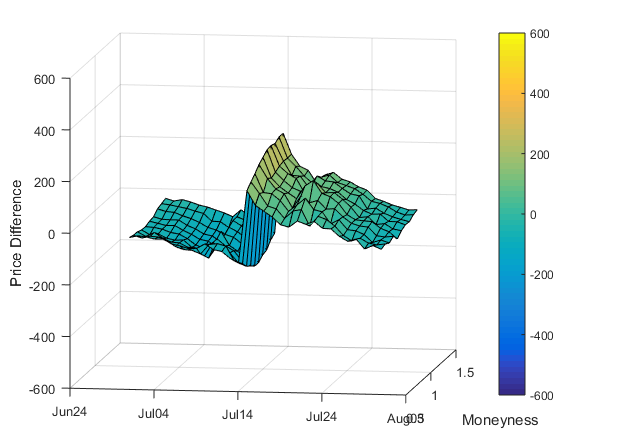}
  \includegraphics[width=8cm,height=5.5cm]{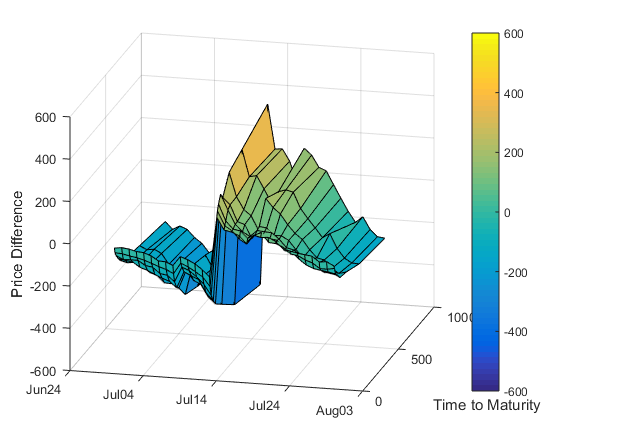}
\footnotesize{\textit{Notes}: This figure plots the option price differences between the SVCJ and SVJ models for July 2017. When looking at moneyness, the time to maturity is fixed at 30 days, and when looking at the time to maturity,  moneyness is ATM. The colour in the graph represents the price difference level; the brighter the colour, the larger the difference between the price from the SVCJ and SVJ models. }
  \label{fig:money1}
\end{subfigure}%
%\begin{subfigure}
%  \centering
 % \includegraphics[width=8cm,height=6cm]{figures/TimetoMaturity_Di%ff}
%  \includegraphics[width=8cm,height=6cm]{figures/TimetoMaturity_Di%ff_bit}
 % \caption{Price difference between the SVJ and SVCJ model plotted %against time to maturity.  CRIX (left panel) and BTC (right %panel).}
%  \label{fig:ttoMaturity1}
%\end{subfigure}
%\caption{plots of....}
%\label{fig:fig}
\end{figure}

Figure \ref{fig:money1} %and \ref{fig:ttoMaturity1} 
displays the estimated BTC call option price differences between the SVCJ and SVJ models with respect to changes in moneyness and across time to maturity for July 2017. It is not hard to see that the pattern is similar to the fitted volatility shown in Figure \ref{fig:vol}. The difference between the SVCJ and the SVJ model is similar besides on July 15 when there is a large spike in the estimated volatility. Therefore, the price differences between the SVCJ and SVJ models are mainly caused by the jumps in the volatility process and the volatility level, which reflects the necessity of adopting the SVCJ model in practice.

%\begin{figure}
%	\begin{center}
	%\caption{BTC call option prices across moneyness and time to maturity}
	%	%\includegraphics[width=6cm,height=6cm]{figures/20170717_Price}
%   \includegraphics[width=10cm,height=9cm]{figures/20170717_price_bit_3_1}\label{fig:price}
     %  \end{center}
%		\footnotesize{\textit{Notes}: This figure plots the call option prices on July 17, 2017 approximated using the parameters reported in Table 4 for SVCJ, SVJ, and SV models. The upper figure shows the call options across different strike prices given time to maturity of 30 days.  The lower figure plots the option prices across different time to maturities given the strike price $K$ of 2250 (at the money). The blue/red/green colored line plots the call option prices from SVCJ/SVJ/SV models.}
		%\hspace*{\fill} \raisebox{-1pt}{\includegraphics[scale=0.05]{qletlogo}\ econ\_SVCJ}
%\end{figure}

\begin{table}[]
\scriptsize{}
%\centering
\caption{Call option price of BTC on $17/07/2017$ from the SVCJ model}
\label{optbitsvcj}
\begin{tabular}{l|llllllll}
\hline\hline
$K \backslash \tau$    & 1    & 7    & 30   & 60  & 90   & 180  & 360  & 720  \\
\hline
1250.00 & 1069.18 & 1017.81 & 1099.87 & 1125.90 & 1157.95 & 1248.98 & 1361.04 & 1365.96 \\
1350.00 & 959.02  & 959.02  & 1006.02 & 1066.67 & 1094.08 & 1224.48 & 1302.60 & 1316.03 \\
1450.00 & 885.20  & 860.15  & 929.32  & 995.45  & 1046.89 & 1099.35 & 1258.83 & 1438.90 \\
1550.00 & 802.38  & 791.34  & 901.27  & 950.34  & 1015.76 & 1114.94 & 1192.24 & 1332.08 \\
1650.00 & 707.97  & 739.10  & 825.07  & 882.17  & 902.32  & 1062.17 & 1175.59 & 1282.36 \\
1750.00 & 625.86  & 678.22  & 786.88  & 856.72  & 896.56  & 962.79  & 1192.61 & 1338.49 \\
1850.00 & 552.26  & 618.94  & 697.11  & 785.62  & 862.83  & 897.74  & 1110.36 & 1289.51 \\
1950.00 & 502.28  & 545.58  & 663.47  & 740.32  & 819.72  & 903.60  & 1052.09 & 1229.45 \\
2050.00 & 425.46  & 511.28  & 629.14  & 741.65  & 772.51  & 905.30  & 1027.76 & 1193.43 \\
2150.00 & 358.30  & 460.57  & 597.44  & 683.55  & 740.64  & 870.66  & 1036.76 & 1164.23 \\
2250.00 & 302.88  & 408.62  & 543.02  & 633.31  & 720.57  & 872.42  & 938.68  & 1051.71 \\
2350.00 & 265.91  & 378.10  & 492.86  & 594.01  & 651.03  & 783.37  & 887.62  & 1064.33 \\
2450.00 & 211.26  & 347.79  & 470.85  & 580.30  & 657.43  & 761.39  & 940.90  & 1085.75 \\
2550.00 & 193.69  & 304.13  & 437.06  & 547.15  & 608.36  & 766.19  & 914.62  & 1101.72 \\
2650.00 & 156.38  & 266.64  & 421.86  & 518.27  & 571.42  & 719.92  & 827.17  & 992.20  \\
2750.00 & 136.24  & 247.38  & 397.92  & 484.70  & 556.31  & 651.86  & 863.10  & 1066.75 \\
2850.00 & 135.28  & 228.47  & 345.42  & 465.75  & 541.61  & 672.76  & 788.25  & 955.97  \\
2950.00 & 100.02  & 202.57  & 341.11  & 413.75  & 488.15  & 627.52  & 780.53  & 917.27  \\
3050.00 & 103.45  & 179.93  & 313.83  & 424.23  & 496.05  & 619.88  & 758.99  & 911.33  \\
3150.00 & 82.59   & 162.72  & 290.90  & 371.20  & 450.85  & 593.10  & 752.88  & 888.89  \\
3250.00 & 72.93   & 140.40  & 273.97  & 358.26  & 442.91  & 571.96  & 726.49  & 933.57\\
\hline\hline
\end{tabular}
\begin{tablenotes}
\item \textit{Notes}: This table reports the approximated call option prices at different time to maturity $\tau$ and strike prices $K$ the SVCJ model on 17/07/2017 based on the parameters reported in Table 1. The numbers in the first row are the time to maturity. The numbers in the first column are the strike prices. The spot BTC price is assumed to be 2250.  
\end{tablenotes}
\end{table}

\subsection{BTC implied volatility smiles}
It is well known that stochastic volatility determines excess kurtosis in the conditional distribution of returns. The excess kurtosis causes symmetrically higher implied Black Scholes volatility when strikes are away from the current prices, e.g., the  level of moneyness is away from the ATM level. This phenomenon is called the "volatility smile". It is well documented in the existing literature that the effect is stronger for short and medium maturity options than for long maturity options for which the conditional returns are closer to normal (\cite{DasSundaram1999}). 
The presence of co-jumps, and the negative correlation between the presence of co-jumps sizes yield additional sources of skewness in the conditional distribution of stock returns (\cite{bandi2016price}). 

To further examine the option-pricing property of BTC, we approximate the implied Black Scholes volatility from various models for different degrees of moneyness (strike/spot) and different times to maturity. First, the European call option prices are simulated using the model parameters reported in Table \ref{tab:parameterbit} for the SVCJ, SVJ and SV models and Table \ref{tab:parameterbitBR} for the BR model. Then the volatility from various models is implied from the Black Scholes model based on the options approximated from different models. We consider four times to maturity: one week, one month, three months and one year. We report the implied volatility surface as a function of moneyness and time to maturity. The results indicate that jumps in returns and volatility include important differences in the shape of the implied volatility (IV) curves, especially for the short maturities options. 

\begin{figure}
\begin{center}
\caption{The IV for the BTC market: the SVCJ, SVJ and SV models}\label{fig:IV_SVCJ}
  \includegraphics[width=12cm,height=9cm]{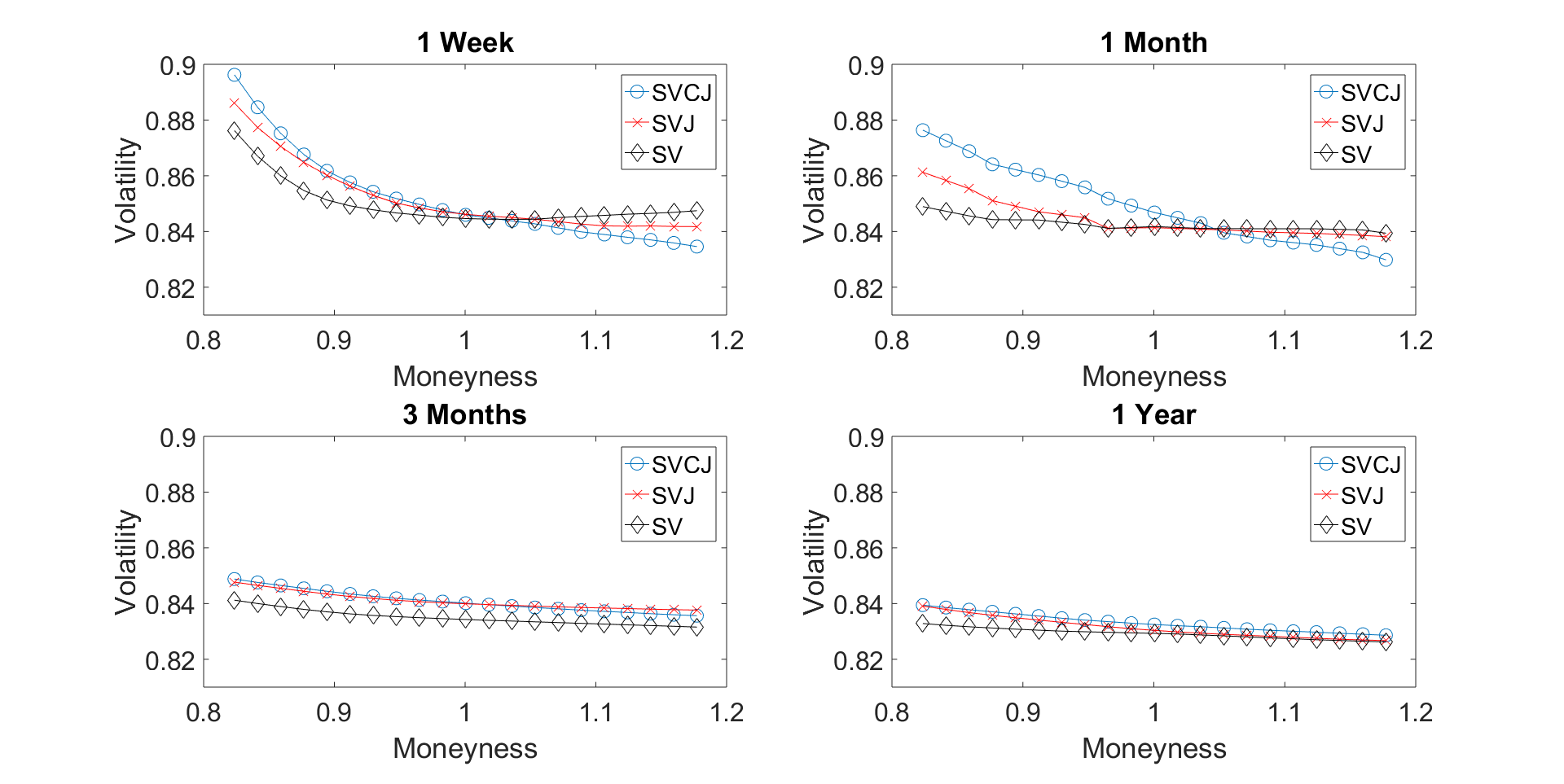}
\end{center}

\footnotesize{
\textit{Notes}: This figure plots the Black Scholes IV for the BTC market based on the SVCJ, SVJ and SV models. The x-axis shows moneyness and the y-axis shows the IV. Four times to maturity have been considered: one week, one month, three months and one year. The lines with 
$\circ$, $\ast$, $\diamond$ plots the IVs of the SVCJ, SVJ and SV models, respectively. }
%\footnote{This figure}
\end{figure}

Figure \ref{fig:IV_SVCJ} shows the IV curves for the SVCJ, SVJ and SV models for four different maturities and across moneyness.  It can be seen from Figure \ref{fig:IV_SVCJ}, that adding jumps in returns steepens the slope of the IV curves. Jumps in volatility further steepen the IV curves. For short maturity options, the difference between the SVCJ, SVJ and SV models for far ITM options is quite large, with the SVCJ model giving the sharpest skewness among the three models. The difference between the SVCJ and SV volatility is approximately 2-3\% for up to one month.  All three models have a one-side volatility skewness. This could be due to the skewness in the conditional distribution of BTC returns (\cite{DasSundaram1999}) and/or that the negative co-jump size yields an additional source of skewness (\cite{bandi2016price}). As time to maturity increases, the volatility curve flattens for all models. According to \cite{DasSundaram1999}, jumps in returns result in a discrete mixture of normal distributions for returns, which easily generates unconditional and conditional non-normalities over short frequencies such as daily or weekly. Over longer intervals, e.g., more than a month, a central-limit effect results in decreases in the amount of excess and kurtosis. Indeed, diffusive stochastic volatility models may generate very flat curves, such as a flat BTC IV for the three-month and the one-year times to maturity. 

However, for the SVCJ model, the curve flattens at a slightly higher level. The implied volatility of the SVJ model is closer to the SVCJ model than the SV model. The difference between the SVCJ, SVJ and SV models becomes larger with short time to maturity options, i.e., the one-week and one-month times to maturity. Similar results have been documented in other studies in which these models have been applied to equity index data. \cite{eraker2003impact}, \cite{eraker2004stock} and \cite{duffie2000transform} find that jumps in returns and variance are important in capturing systematic variations in Black- Scholes volatility. 
In general, although the BTC market has the unique feature of having more jumps, which makes it different from other mature markets (e.g., equity), the option prices and the IV from the affine models generally follow the conventional characteristics reported from other option markets. 

\begin{figure}
\begin{center}
\caption{The IV for the BTC market: BR model}\label{fig:IV_BR}
  \includegraphics[width=16cm,height=13cm]{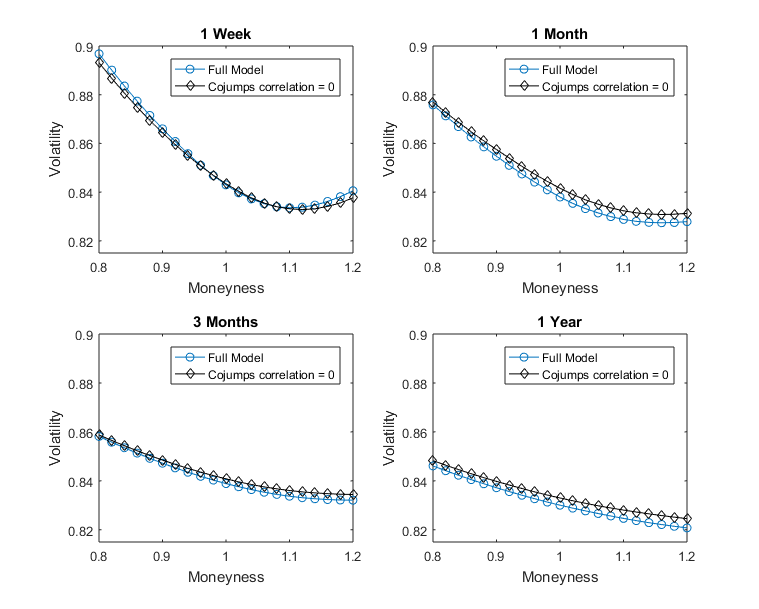}
\end{center}

\footnotesize{\textit{Notes}: This figure plots the implied Black Scholes volatility for the BTC option prices based on the BR model. The x-axis shows moneyness, and the y-axis shows the IV. Four times to maturity have been considered: one week, one month, three months and one year. The IVs are based on the simulated option prices using the model parameters reported in Table {2}. The full model uses parameters from column 4 of Table {2}. A co-jumps correlation of 0 means that $\rho_J$ is set to zero while the other parameters remain the same as in the full model. }
\label{IV_Bandi}
\end{figure}

We have also estimated the BR IVs with the same time to maturity and moneyness used for the SVCJ IVs. We simulate the option prices using the model parameters reported in column 4 of Table \ref{tab:parameterbitBR}. We distinguish the case of $\rho_J$, which is set to be a model-fitted parameter from the SVCJ fit or to be zero, i.e., the IV surface corresponds to a case with a correlation between jump sizes equaling -0.5257 or a correlation between jump sizes equaling to zero. The IVs as a function of moneyness from the BR model are plotted in Figure \ref{fig:IV_BR}. We can see that the IVs of the BR model agree with the SVCJ model. We see a one-side volatility skewness, i.e., the ITM call option prices are higher than the OTM call options. However, due to the significantly negative jump-size correlation $\rho_J$, the slope of the IVs from the BR full model is steeper than the BR model with a case of uncorrelated jump sizes. The impact of the negative jump size correlation is stronger for short time to maturity options, i.e., the one-week and one-month times to maturity.  This is mentioned in the results of \cite{duffie2000transform} as well, who find a superior fit of the IV smirk when calibrating a more negative correlation between jump sizes. Similarly, \cite{eraker2004stock} finds a statistically significant correlation between jump size only when employing option data in addition to returns data. \cite{bandi2016price} also shows that anti-correlated jump sizes are a fundamental property of prices and volatility. However, the use of high-frequency data is sufficient to reveal this property with no further need for option data. 

%\ref{fig:IV_Bandi} 

\section{The CRyptocurrency IndeX (CRIX)}\label{sec:crix}
The CRyptocurrency IndeX, a value-weighted cryptocurrency market index with an endogenously determined number of constituents using some statistical criteria, is described in \cite{hardle2015crix} and further sharpened in \cite{trimborn2018crix}. It is constructed to track the entire cryptocurrency market performance as closely as possible. The representativity and the tracking performance can be assured as CRIX considers a frequently changing market structure. The reallocation of the CRIX happens on a monthly and quarterly basis (see \cite{trimborn2018crix} and {\tt thecrix.de} for details). CRIX has been widely investigated in the pioneering research on cryptocurrencies, including by \cite{chen2017first}, \cite{Hafner2018}, \cite{chen2019sentiment} and \cite{da2019herding}. 

There are two advantages of holding a portfolio comprising a wide variety of Cryptocurrencies like CRIX. The first advantage is the diversification benefit. The evidence from \cite{hardle2019understanding} shows that the correlations among the most leading coins are around 0.5, indicating a promising potential of diversification. The correlations among coins vary over time, as shown in \cite{hardle2019understanding}. It shows that the diversification effect through forming a portfolio is beneficial, although this effect may vary over time. 

The second advantage underscores that the efficient portfolio, like CRIX, entails a higher Sharpe ratio than that of BTC. From the view of {institutional} investors, a smart strategy is to hold a market portfolio comprising of the coins with sufficient liquidity and market capitalization to leverage between profitability and risk-sharing. A simple calculation of the annual Sharpe ratio for both BTC and CRIX-based portfolios sheds some light. The Sharpe ratios of CRIX in 2016 and 2017 are respectively 0.094 and 0.194, however, the ratios of BTC are relatively lower (0.085 in 2016 and 0.149 in 2017). It suggests that investors should rather look at all possible portfolios in an investment opportunity set that potentially optimize their mean-variance preference.

Given the merits of portfolio deployment over a single altcoin investment rule, institutional investors may demand the corresponding derivatives for hedging position risk. The options with a cryptocurrency index (CRIX) as underlying may fulfill such needs in practice. Apart from hedging purposes, and for speculators without any position, such index options are quite precious and enable them to bet on future movement.

{Therefore w}e perform an analysis for CRIX. {All econometric models have been estimated with the CRIX data.} We summarize our major findings here and place the supplementary parts in the appendix. In brief, all the model parameters estimated with CRIX convey a similar configuration as estimated with BTC, e.g., the mean jump size of the CRIX volatility process reported in Table \ref{tab:parameterCRIX} is 0.709, which is 0.620 {for BTC shown in Table \ref{tab:parameterbit}. The estimated volatility from the SVCJ and SVJ models (see Figure \ref{fig:jump2_crix}) shows that the jumps are better captured by the SVCJ than the SVJ model.  In addition,  {Figure \ref{fig:Option_CRIX}} displays the call option prices surface contour plot from the SVJ, SV and SVCJ models with respect to changes in moneyness and  time to maturity. It shows that the SVCJ model has more volatile patterns than those of the SVJ and SV models with the BTC options.} In general, we confirm the consistency between BTC and the CRIX.

\newpage
\section{Conclusion}\label{sec:con}

"The Internet is among the few things that humans have built that they do not truly understand" according to \cite{cohen2017}. Cryptocurrency, a kind of innovative internet-based asset, brings new challenges but also new ways of thinking for economists, cliometricians and financial specialists. Unlike classic financial markets, the BTC market has a unique market microstructure created by a set of opaque, unregulated, decentralized and highly speculation driven markets.

This study provides a way of pricing cryptocurrency derivatives using advanced option-pricing models such as the SVCJ and BR models. We find that in general, the SVCJ model performs as well as the non-affine BR model. We especially find that the correlation between the jump sizes in returns and the volatility process is anti-correlated. The jump-size correlation is statistically (marginally) negative in the BR (SVCJ) model. Deviating from the equity market, we cannot obtain a  significant negative "leverage effect" parameter $\rho$, which implies a nonnegative relation between returns and volatility.  The reason for this relationship might be that BTC is different from the conventional stock market, not only because the BTC market is highly unregulated but also due to the fact that the BTC price is not informative (as there are no fundamentals allowing the BTC market to set a "fair" price) and is driven by emotion and sentiment. This speculative behaviour can be explained by the  "noise trader" theory from \cite{Kyle1985}. The positive relation might result from the fact that BTC investors irrationally act on noise as if it were information that would give them an edge. 

%The results from our option pricing indicate that the option price of BTC obtains the consistent property observed in equity option market. We find that the option prices increase in moneyness from ITM to Out-Of-Money OTM and increase in time to maturity. The option prices are increasing with the time to maturity and with the increases in volatility level.  The results from the IV indicate that adding jumps in returns steepens the slope of the implied volatility (IV) curves. The further steep IV curve can be reinforced by the presence of jumps in volatility. The presence of co-jumps and the anti-correlation between the jump size in return and variance steepen the IV smile curves furthermore. The short time to maturity is more sensitive to jumps and anti-correlated jump sizes.

We find that option prices are very much driven by jumps in the returns and volatility processes and co-jumps between the returns and volatility.
This can be seen from the shape of the IV curves.
This study provides a grounding base, or an anchor, for future studies which aim to price cryptocurrency derivatives. This study provides useful information for establishing an options market for BTC in the near future.
\newpage

\newpage

\section{Appendix}\label{sec:app}
We provide preliminary fit results of econometric models on the Bitcoin time series. We also collect results on analysis of the CRIX.
\subsection{ARIMA}\label{sec:arima}

\begin{figure}
\begin{center}
\caption {ACF and PACF of BTC} \label{fig:acf}
\begin{tabular}{cc}
\includegraphics[width=12cm,height=8cm]{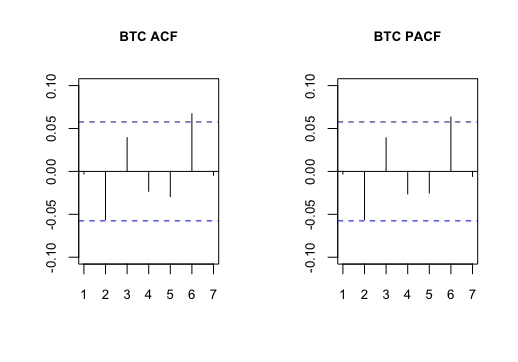}
\end{tabular}
\vspace{-0.5cm}
\end{center}
	\footnotesize{\textit{Notes}: This figure plots the ACF and PACF for BTC returns. The returns are the log-first difference calculated based on the price from 01/08/2014 to 29/09/2017.  The x-axis plots the lags, and the y-axis plots the ACF and PACF values.}
\end{figure}

%Figure \ref{fig:BTCprice} indicates that BTC prices do not behave like conventional stock prices. One records extremely high volatility and scattered spikes. These prices are far from being stationary. % The plot of returns shows that volatility is time-varying and clustering. Therefore we start to fit the returns of BTC with the GRACH model.
We first fit an ARIMA model. After an inspection through the ACF and PACF plot in Figure \ref{fig:acf}, we start with an ARIMA($p, d, q$) model,
\begin{equation}
       a(L)\Delta y_{t}= b_{L}\varepsilon_{t}
\end{equation}
where $y_{t}$ is the variable of interest, $\Delta y_{t}= y_{t}- y_{t-1}$, $L$ is the lag operator and $\varepsilon_{t}$ a stationary error term. Model selection criteria such as AIC or BIC indicates that the ARIMA($2, 0, 2$) is the model of choice. The parameters estimated from the ARIMA(2,0,2) are reported in Table \ref{tab:arima}. The significant negative signs in $a_1$ and $a_2$ indicate an overreaction, that is, a promising positive return today leads to a return reversal in the following two days or vice versa. Hence, the CC markets tend to overreact to good or bad news, and this overreaction can be corrected in the following two days. An ARIMA model for the CC assets, therefore, suggests predictability due to an ``overreaction''.
The Ljung-Box test confirms that there is no serial dependence in the residuals based on the ARIMA($2, 0, 2$) specification. Note that the squared residuals carry incremental information that is addressed in the following GARCH analysis.
\begin{table}
 \begin{threeparttable}
% \begin{center}
 %\begin{small}
\caption{Estimation result of ARIMA(2,0,2)}\label{tab:arima}
\begin{tabular}{{p{2.5cm}p{0.5cm}p{2.8cm}p{0.5cm}p{3.5cm}}}
\hline\hline
 &  &{Bitcoin}&& \\
\hline
{Coefficients} &&  {Estimate} && Standard error (Robust)  \\
\hline
intercept $c$ & & 0.002 && 0.001 \\
$a_{1}$       &  &  -0.867&&0.304 \\
$a_{2}$        &  & -0.596&&0.177 \\
$b_{1}$     & &0.868 & &0.321 \\
$b_{2}$      & &0.539& & 0.190 \\
\hline\hline
\end{tabular}
%\end{small}
\hspace{-0.1cm}
%\centering
%\begin{tablenotes}
\footnotesize
\textit{Notes}: This table reports the parameter estimated from ARIMA (2,0,2) with BTC daily returns. The residual distributions are assumed to be Gaussian. The maximized likelihood value is 2231.7. The AIC and BIC are -4451.4 and -4415.74, respectively.     %\end{tablenotes}
%\end{center}
 \end{threeparttable}
  \end{table}

% \begin{table}\label{arima}
% %\begin{small}
% \begin{center}
% \begin{tabular}{c r@{.}l r@{.}l r@{.}l r@{.}l}
% \hline\hline
%  &  \multicolumn{4}{c}{CRIX} &   \multicolumn{4}{c}{Bitcoin}\\
% \hline
%  Coefficients &  \multicolumn{2}{c}{Estimate} &  \multicolumn{2}{c}{Standard deviation} &  \multicolumn{2}{c}{Estimate} &  \multicolumn{2}{c}{Standard deviation} \\
% \hline
% intercept $c$ & \enspace 0&002 & \enspace 0&001 & \enspace -0&002 & \enspace 0&001 \\
% $a_{1}$  & \enspace -0&819 & \enspace 0&188     & \enspace -0&521 & \enspace 0&159 \\
% $a_{2}$  & \enspace -0&791 &  \enspace 0&112    & \enspace -0&747 & \enspace 0&160 \\
% $b_{1}$  &  \enspace 0&828 &  \enspace 0&207    & \enspace 0&467 & \enspace 0&168 \\
% $b_{2}$ & \enspace 0&746 & \enspace0&127        & \enspace 0&700 & \enspace 0&176 \\
% Log lik & \enspace 2243&360 & \enspace& & \enspace 2139&340 & \enspace&  \\
% \hline\hline
% \end{tabular}
% \caption{Estimation result of ARIMA(2,0,2) model}
% \end{center}
% %\end{small}
% \end{table}

\subsection{GARCH Model}\label{sec:garch}
The GARCH model, introduced first by \cite{bollerslev1986generalized}, reflects the changes in the conditional volatility of the underlying asset in a parsimonious way. The volatility properties of digital currency assets have been studied in a vast amount of literature that applies GARCH-type methods \citep{hotz2018predicting, chu2017garch, chan2017statistical, conrad2018long}.

Let us start with a GARCH-type model for characterizing the conditional variance process of BTC. The ARIMA-$t$-GARCH model with $t$-distributed innovations used to capture fat tails is as follows:
\begin{eqnarray}
       a(L)\Delta y_{t}&=& b_{L}\varepsilon_{t}  \\
       \varepsilon_{t} &=& Z_{t}\sigma_{t}, \quad Z_{t} \sim t(\nu) \nonumber \\
        \sigma_{t}^{2} &=& \omega  + \beta_{1}\sigma_{t-1}^{2} + \alpha_{1}\varepsilon_{t-1}^{2}
        \label{equ:tgarch}
\end{eqnarray}
where $\sigma_{t}^{2}$ represents the conditional variance of the process at time $t$ and $t(\nu)$ refers to the zero-mean $t$ distribution with
$\nu$ degrees of freedom. The choice of the $t$-distribution rather than the Gaussian distribution is supported by \cite{hotz2018predicting} and \cite{chan2017statistical}.

The covariance stationarity constraint $\alpha_{1}+\beta_{1} < 1$ is imposed. As shown in Table \ref{tab:tgarch}, the $\beta_{1}$ estimate from BTC indicates a persistence in the variance process, but its value is relatively smaller than those estimated from the stock index returns (see \cite{franke2019statistics}). Typically, the persistence-of-volatility estimates are very near to one, showing that conditional models for stock index returns are very close to being integrated. By comparison, BTC places a relatively higher weight on the $\alpha_{1}$ coefficient and relatively lower weight on the $\beta_{1}$ to imply a less-smooth volatility process and striking disturbances from the innovation term. This may further imply that the innovation is not pure white noise and can occasionally be contaminated by the presence of jumps.

In addition to the property of leptokurtosis, the leverage effect is commonly observed in practice. According to a large body of literature, starting with \cite{engle1993measuring}, the leverage effect refers to an asymmetric volatility response given a negative or positive shock. The leverage effect is captured by the exponential GARCH (EGARCH) model by \cite{nelson1991conditional},
\begin{eqnarray}
\varepsilon_{t} &=& Z_{t}\sigma_{t} \nonumber \\
Z_{t} &\sim & t(\nu) \nonumber \\
\log (\sigma_{t}^{2} )&=& \omega  + \sum_{i=1}^{p}\beta_{i} \log ( \sigma_{t-i}^{2}) + \sum_{j=1}^{q} g_{j} \left(Z_{t-j}\right) \label{equ:egarch}
\end{eqnarray}
where $g_{j} \left(Z_{t}\right) = \alpha_{j} Z_{t} + \phi_{j}( \lvert Z_{t-j}\rvert - {\mathop{\mbox{\sf E}}}\lvert Z_{t-j}\rvert)$ with $j=1,2,\ldots,q$. When $\phi_{j} = 0$, we have the logarithmic GARCH (LGARCH) model from \cite{geweke1986modelling} and \cite{pantula1986comment}. To accommodate the asymmetric relation between stock returns and volatility changes, the value of $g_{j} \left(Z_{t}\right)$ must be a function of the magnitude and the sign of $Z_{t}$. Over the range of $0< Z_t < \infty$, $g_{j} \left(Z_{t}\right)$ is linear in $Z_{t}$ with slope $\alpha_{j} + \phi_{j}$, and over the range $-\infty < Z_t \leq 0$, $g_{j} \left(Z_{t}\right)$ is linear in $Z_{t}$ with slope $\alpha_{j} - \phi_{j}$.

The estimation results based on the ARIMA(2,0,2)-$t$-EGARCH(1,1) model are reported in Table \ref{tab:egarch}. The estimated $\alpha_{1}$ is no longer significant, showing a vanished sign effect. However, a significant positive value of $\phi_{1}$ indicates that the magnitude effect represented by $\phi_{1}( \lvert Z_{t-1}\rvert - {\mathop{\mbox{\sf E}}}\lvert Z_{t-1}\rvert)$ plays a bigger role in the innovation in $\log (\sigma_{t}^{2} )$. %{\red The presence of jumps might be one of the causes of the magnitude effect (need to be confirmed!)}.

\begin{table}
\begin{threeparttable}
%\begin{center}
\caption{Estimated coefficients of $t$-GARCH(1,1)}\label{tab:tgarch}
%\begin{tabular}{cccr}
\begin{tabular}{p{2cm}p{0.5cm}p{2.5cm}p{0.5cm}p{2.5cm}p{0.2cm}p{2cm}}
\hline\hline
 Coefficients && Estimates && Robust std && $t$ value\\
\hline
BTC && && && \\
\hline
$\omega$ && $3.92e-05$ &&  $1.49e-05$ &&   $\enspace 2.63$ \\
$\alpha_{1}$ && $2.28e-01$ & & $4.46e-02$  &&  $\enspace 5.12$ \\
$\beta_{1}$ && $7.70e-01$ && $5.13e-02 $ & & $\enspace 14.98$ \\
$\nu$ & &$3.64e+00$ &&    $4.08e-01$ & &  $\enspace 8.91$\\
% \hline
% CRIX & & & \\
% \hline
% $\omega$ & $4.93e-05$ &  $2.69e-05$ &   $\enspace 1.83$ \\
% $\alpha_{1}$ & $2.23e-01$  & $4.28e-02$  &  $\enspace 5.45$ \\
% $\beta_{1}$ & $7.76e-01$ & $5.62e-02 $ &  $\enspace 13.81$ \\
% $\nu$ & $3.10e+00$ &    $2.19e-01$ &   $\enspace 14.15$\\
\hline\hline
\end{tabular}
%\begin{tablenotes}
\vspace{0.1in}

\footnotesize
\textit{Notes}: This table reports the estimated parameters from the t-GARCH(1,1) model. The robust version of standard errors (robust std) are based on the method of \cite{white1982maximum}.
%\end{tablenotes}
%\hspace*{\fill} \raisebox{-1pt}{\includegraphics[scale=0.05]{qletlogo}\
%\href{https://github.com/QuantLet/EconCrix/tree/master/econ-tgarch}{econ\_tgarch} }
%\end{center}
\end{threeparttable}
\end{table}

\begin{table}
\begin{threeparttable}
%\begin{small}
\caption{Estimated coefficients of $t$-EGARCH(1,1) model} \label{tab:egarch}
%\begin{center}
\begin{tabular}{p{2cm}p{0.5cm}p{2.5cm}p{0.5cm}p{2.5cm}p{0.2cm}p{2cm}}
\hline\hline
 Coefficients && Estimates && Robust std && $t$ value \\
\hline
BTC && && && \\
\hline
$\omega$ & &$3.84e-05$ &&  $1.47e-05$ & &  $\enspace 2.61$ \\
$\alpha_{1}$ & &$1.05e-03$  && $5.10e-02$  &&  $\enspace 0.98$ \\
$\beta_{1}$ && $9.52e-01$ && $1.54e-02 $ &&  $\enspace 61.73$ \\
$\phi_{1}$ && $4.16e-01$ &&    $6.64e-02$ & &  $\enspace 6.25$\\
$\nu$ && $3.26e+00$ & &   $4.16e-01$ & &     $\enspace 7.82$\\
% \hline
% CRIX & & & \\
% \hline
% $\omega$ & $4.93e-05$ &  $2.69e-05$ &   $\enspace 1.83$ \\
% $\alpha_{1}$ & $5.58e-02$  & $4.34e-02$  &  $\enspace 1.34$ \\
% $\beta_{1}$ & $9.62e-01$ & $1.38e-02 $ &  $\enspace 69.43$ \\
% $\phi_{1}$ & $5.36e-01$ &    $1.39e-01$ &   $\enspace 3.85$\\
% $\nu$      & $2.42e+00$ &    $2.42e-01$ &   $\enspace 10.02$\\
\hline\hline
\end{tabular}
%\begin{tablenotes}\footnotesize
%\item[*] The robust version of standard errors (robust std) are based on the method of \cite{white1982maximum}.
%\end{tablenotes}
%\hspace*{\fill} %\raisebox{-1pt}{\includegraphics[scale=0.05]{qletlogo}\
 %\href{https://github.com/QuantLet/EconCrix/tree/master/econ-garch}{econ\_garch} }
\footnotesize
\textit{Notes}: This table reports the estimated parameters from the t-EGARCH(1,1) model. The robust version of standard errors (robust std) are based on the method of \cite{white1982maximum}.
%\end{small}
%\end{center}
\end{threeparttable}
\end{table}

\begin{figure}[htp]
\begin{center}
\caption {The QQ plot for BTC based on the residuals of $t$-GARCH(1,1) model} \label{fig:11qqstu}
\begin{tabular}{c}
\includegraphics[width=8cm,height=6cm]{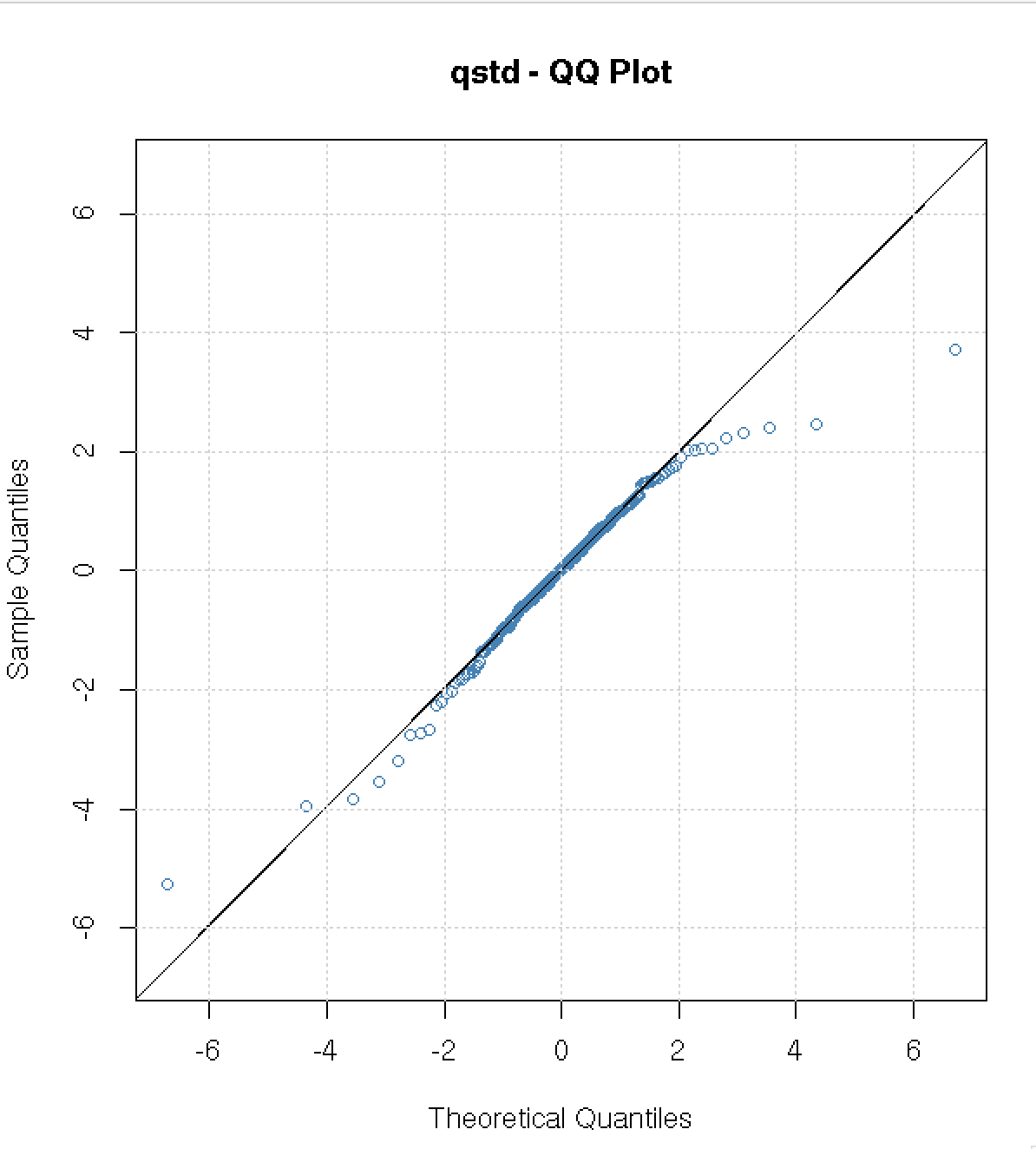}
\end{tabular}
\vspace{-0.5cm}
\end{center}
\end{figure}

We compare the model performances between two types of GARCH models through information criteria, and a $t$-EGARCH(1,1) model is suggested.
Note that, as shown in Figure \ref{fig:11qqstu}, the QQ plots demonstrate a deviation from the student-$t$. In \cite{chen2017first}, GARCH and variants such as $t$-GARCH, EGARCH have been reported, and, while they are seen to fit the dynamics of BTC nicely, they still could not handle the extreme tails in the residual distribution. Equipped with these findings and taking into account the occasional interventions, we opt for the models with jumps for better characterization of CC dynamics. The presence of jumps is indeed more likely in this decentralized, unregulated and illiquid market. Numerous political interventions also suggest the introduction of the jump component into a pricing model.

\subsection{CRIX}
\begin{figure}
	\begin{center}
      \includegraphics[width=10cm,height=8cm]{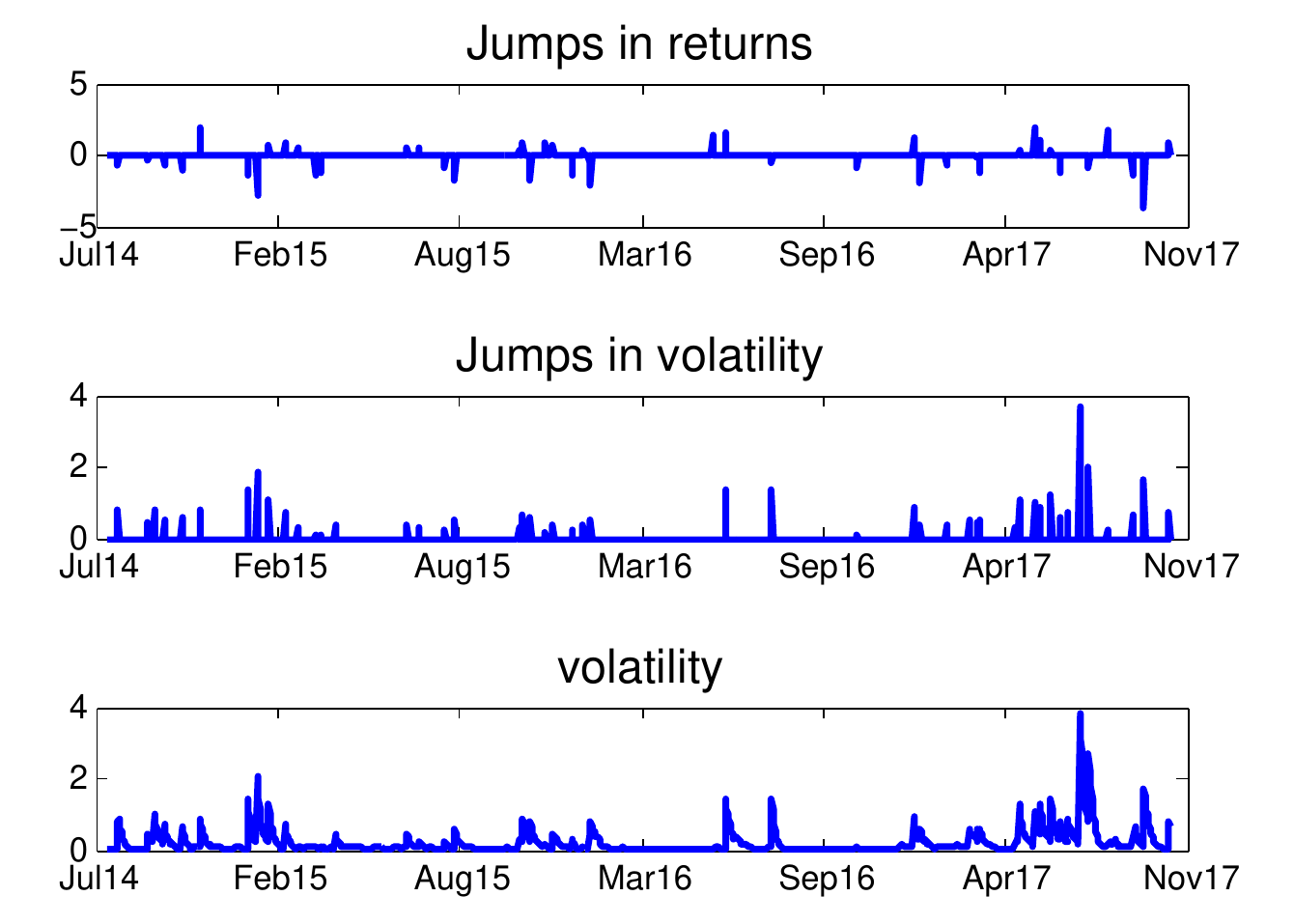}
		\caption{Jumps estimated in returns and volatility from the SVCJ model: CRIX}\label{fig:jump1est_crix}
		%\hspace*{\fill} %\raisebox{-1pt}{\includegraphics[scale=0.0%5]{qletlogo}\ econ\_SVCJ}
	\end{center}
\end{figure}

\begin{figure}
	\begin{center}
\includegraphics[width=10cm,height=7cm]{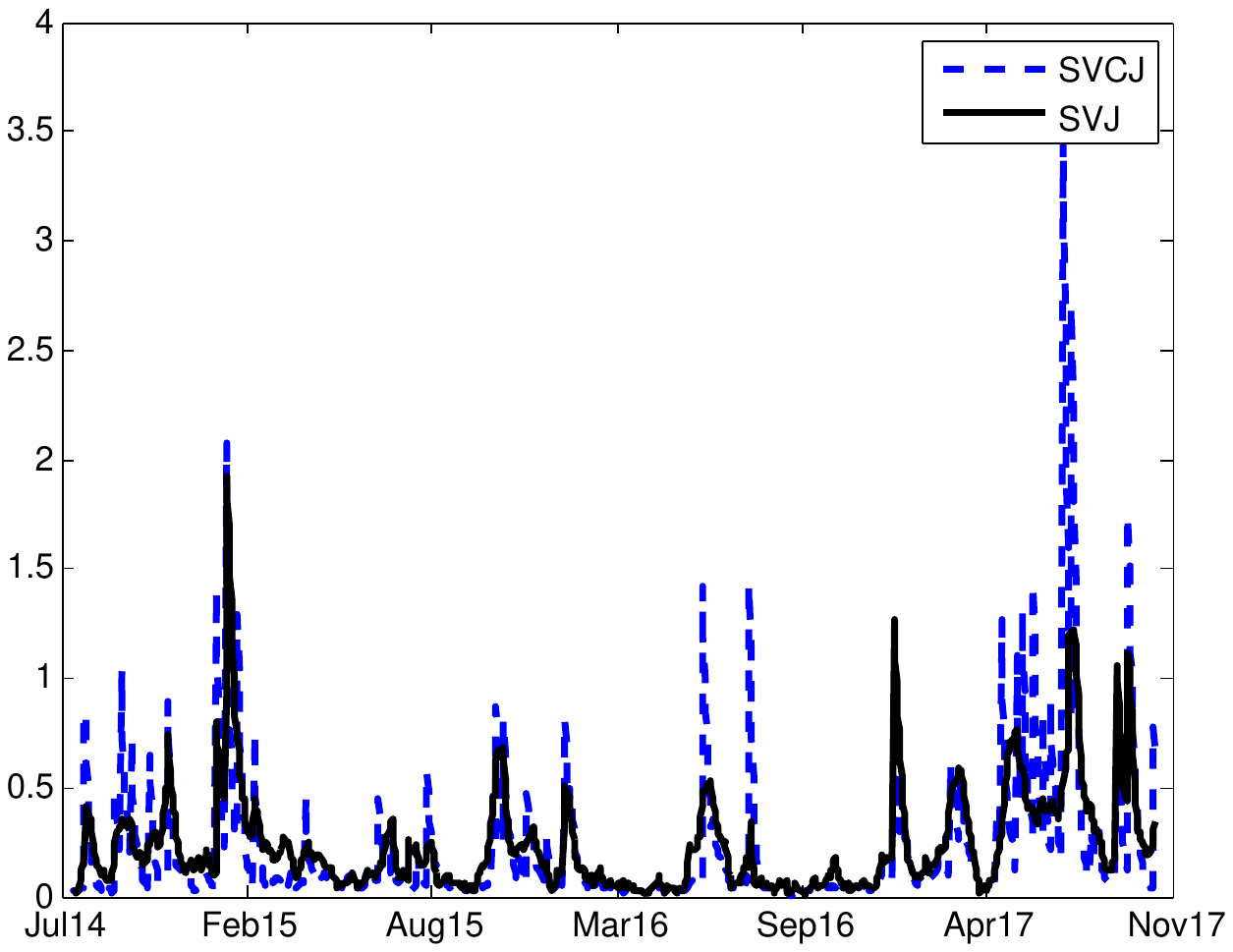}
		\caption{Estimated volatility from the SVCJ and SVJ models: CRIX}\label{fig:jump2_crix}
	%	\hspace*{\fill} \raisebox{-1pt}{\includegraphics[scale=0.05]{qletlogo}\ econ\_SVCJ}
	\end{center}
\end{figure}

\begin{figure}
\begin{subfigure}
  \centering
  \caption{Call option prices across moneyness and time to maturity: CRIX }
   \includegraphics[width=8cm,height=6cm]{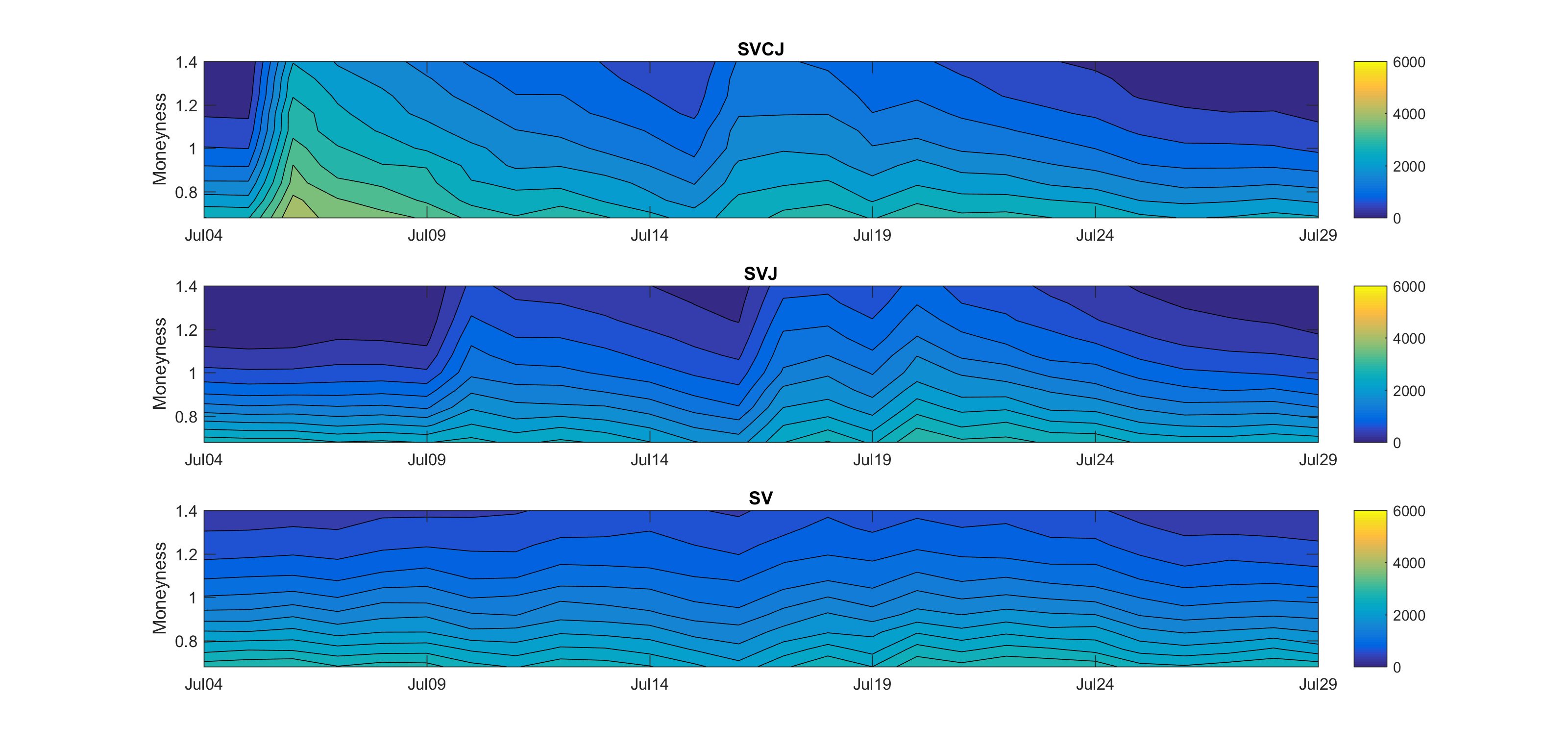}
   \includegraphics[width=8cm,height=6cm]{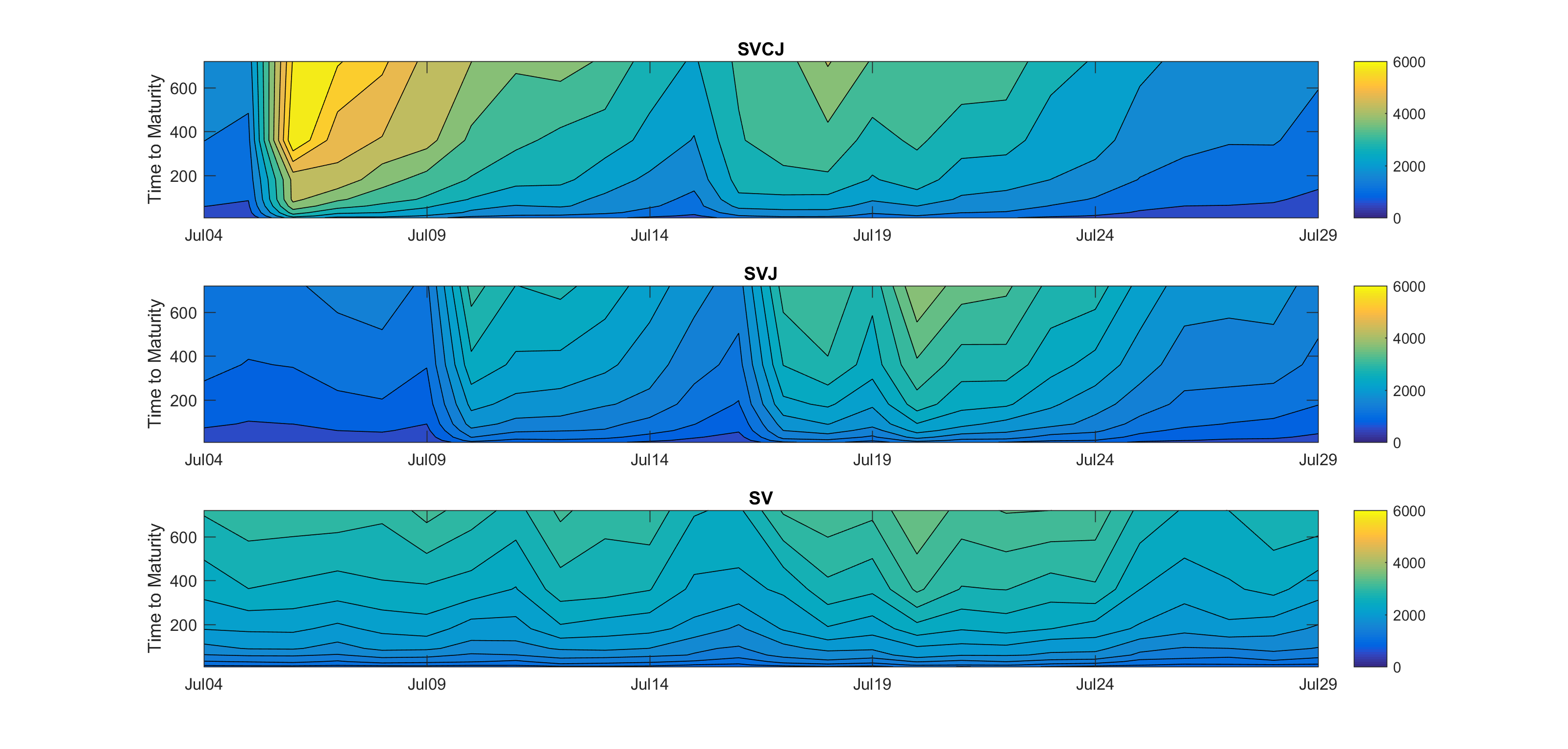}
  \footnotesize{\textit{Notes}: This figure graphs the call option prices surface counterplot across moneyness and time to maturities for the month of July 2017 for CRIX. When looking at moneyness, the time to maturity is fixed at 30 days, and when looking at the time to maturity,  moneyness is ATM. The colour in the graph represents the price level; the brighter the colour, the higher the price. } 
  \label{fig:Option_CRIX}
\end{subfigure}%
%\begin{subfigure}
 % \centering
  %\includegraphics[width=8cm,height=6cm]{figures/TimetoMaturity}
  % \includegraphics[width=8cm,height=6cm]{figures/TimetoMaturity_b%it}
 % \caption{Call option price against time to maturity for BTC %on 201707.}
%  \label{fig:ttoMaturity}
%\end{subfigure}
%\caption{plots of....}
%\label{fig:fig}
\end{figure}

This appendix presents the empirical results of CRIX covering %(1) the econometric analysis of its dynamics shown in Tables \ref{tab:tgarch_crix} and \ref{tab:egarch_crix} 
(1) jumps in returns and volatility from the SVCJ model shown in Figure \ref{fig:jump1est_crix} and (2) the estimated volatility from the SVCJ and SVJ models shown in Figure \ref{fig:jump2_crix}. (3) The estimated call options across moneyness and time to maturity in Figure \ref{fig:Option_CRIX}.  In general, a general consistency can be found between CRIX and BTC. Other results are available upon request.


\begin{table}[h!]
\centering
 \begin{threeparttable}
\caption{Parameters for the SVCJ, SVJ and SV models: CRIX } \label{tab:parameterCRIX}
\scriptsize{}
 \centering
  %\begin{tabular}{P{2.5cm}P{2.5cm}P{2.5cm}}
\begin{tabular}{p{0.9cm}p{2.5cm}p{0.2cm}p{2.5cm} p{0.2cm}p{2.5cm}}
\\ \hline\hline
               & $SVCJ$& & $SVJ$&& SV    \\ [0.1cm]  \hline
  $\mu$  &     0.042   & &    0.0437       & &0.017   \\  
             &[0.030, 0.054]&& [0.027, 0.061] && [0.000	0.034]\\[0.1cm]
$\mu_y$  &    -0.0492    & &-0.515     & &  - \\ 
             &[-0.777, 0.678]&& [-1.110, 0.079] && -\\[0.1cm]
$\sigma_y$& 2.061   & &2.851      & &  - \\ 
             &[1.214, 2.907]&& [1.349, 4.354] && -\\[0.1cm]
$\lambda$  &     0.0515    & &    0.035       & &-   \\ 
             &[0.038, 0.065]&& [0.017, 0.052] && -\\[0.1cm]
$\alpha$    &    0.0102   & &    0.026      & &0.010 \\ 
             &[0.009, 0.012]&& [-0.012	0.063] && [0.007	0.012]\\[0.1cm]
$\beta$     &    -0.188    & &   -0.240     & &-0.038 \\ 
            &[-0.205, -0.170]&& [-0.383, -0.096] && [-0.056	-0.020]\\[0.1cm]
$\rho$       &   0.275    & &  0.214    & &0.003 \\  
            &[0.140, 0.409]&& [0.014, 0.415] && [-0.130	0.136]\\[0.1cm]
$\sigma_v$  &   0.007   & & 0.016    & &0.018 \\  
            &[0.005, 0.009]&& [-0.001, 0.033] && [0.014	0.022]\\[0.1cm]
$\rho_j$ &  -0.210  & &  -    & &- \\  
            &[-0.924, 0.503]&& - && -\\[0.1cm]
$\mu_v$ &  0.709 & &  -    & &- \\  
            &[0.535, 0.883]&& - && -\\[0.1cm]
$MSE$   &0.673	   & &0.707  &&0.736\\
\hline\hline
\end{tabular}
\begin{tablenotes}
 \footnotesize
 \item{\textit{Notes}: The table reports posterior means and 95\% credibility intervals (in square brackets) for the parameters of the SVCJ, SVJ and SV models. All parameters are estimated using CRIX daily returns calculated as the log difference based on the prices from 01/08/2014 to 29/09/2017. }
\end{tablenotes}
\end{threeparttable}
%\end{center}
 \end{table}

% \begin{figure}
% 	\begin{center}
% 	\includegraphics[width=60mm]{figures/qqgarch}
% 	\includegraphics[width=60mm]{figures/qqsv} \\
% 	\includegraphics[width=60mm]{figures/qqsvj}
% 	\includegraphics[width=60mm]{figures/qqsvcj}\\
% 	\caption{Normal probability plots for SVCJ, SVJ, SV models for Crix.} \label{fig:qq}
%   \end{center}
% \end{figure}
% \input{response_ report_JoFE.tex}

\newpage
\bibliographystyle{apalike}
%\addcontentsline{toc}{section}{References}
\bibliography{opbib}
\end{document}